\documentclass[apj]{emulateapj}
\usepackage{amsmath}
\usepackage{threeparttable}
\usepackage{multirow}

\shorttitle{KYDISC: Morphology, Quenching, and Mergers in Clusters}
\shortauthors{Oh et al.}

\def\deg{^\circ}
\def\sur{\rm mag\,\, arcsec^{-2}}
\def\umag{u^{\prime}}
\def\gmag{g^{\prime}}
\def\rmag{r^{\prime}}
\def\imag{i^{\prime}}
\def\zmag{z^{\prime}}

\begin{document}

\title{KYDISC: Galaxy Morphology, Quenching, and Mergers in the Cluster Environment}

\author{Sree Oh$^{1, 2, 3}$ }
\author{Keunho Kim$^{4}$}
\author {Joon Hyeop Lee$^{5, 6}$}
\author {Yun-Kyeong Sheen$^{5}$}
\author {Minjin Kim$^{5,6}$}
\author {Chang H. Ree$^{5}$}
\author {Luis C. Ho$^{7,8}$}
\author {Jaemann Kyeong$^{5}$}
\author {Eon-Chang Sung$^{5}$}
\author {Byeong-Gon Park$^{5, 6}$}
\author{Sukyoung K. Yi$^{3}$ }

\affil{
$^1$ Research School of Astronomy \& Astrophysics, The Australian National University, Canberra, ACT 2611, Australia; sree.oh@anu.edu.au\\
$^2$ ARC Centre of Excellence for All Sky Astrophysics in 3 Dimensions (ASTRO 3D), Australia\\
$^3$ Department of Astronomy \& Yonsei University Observatory, Yonsei University, Seoul 03722, Korea; yi@yonsei.ac.kr\\ 
$^4$ School of Earth \& Space Exploration, Arizona State University, Tempe, AZ 85287, USA \\
$^5$ Korea Astronomy and Space Science Institute, Daejeon 34055, Korea\\
$^6$ University of Science and Technology, Daejeon 34113, Korea\\
$^7$ Kavli Institute for Astronomy and Astrophysics, Peking University, Beijing 100871, China\\
$^8$ Department of Astronomy, School of Physics, Peking University, Beijing 100871, China\\
}

\begin{abstract}
We present the KASI-Yonsei Deep Imaging Survey of Clusters (KYDISC) targeting 14 clusters at $0.015 \lesssim z \lesssim 0.144$ using the Inamori Magellan Areal Camera and Spectrograph on the 6.5-meter Magellan Baade telescope and the MegaCam on the 3.6-meter Canada-France-Hawaii Telescope. We provide a catalog of cluster galaxies that lists magnitudes, redshifts, morphologies, bulge-to-total ratios, and local density. Based on the 1409 spectroscopically-confirmed cluster galaxies brighter than -19.8 in the $r$-band, we study galaxy morphology, color, and visual features generated by galaxy mergers. We see a clear trend between morphological content and cluster velocity dispersion, which was not presented by previous studies using local clusters. Passive spirals are preferentially found in a highly dense region (i.e., cluster center), indicating that they have gone through the environmental quenching. In deep images ($\mu_{\rmag}\sim$ 27 $\sur$), 20\% of our sample show signatures of recent mergers, which is not expected from theoretical predictions and a low frequency of ongoing mergers in our sample ($\sim 4\%$). Such a high fraction of recent mergers in the cluster environment supports a scenario that the merger events that made the features have preceded the galaxy accretion into the cluster environment. We conclude that mergers affect a cluster population mainly through the pre-processing on recently accreted galaxies.
 \end{abstract}

\section{Introduction}
\label{sec:intro}

A galaxy cluster is the most dramatic place to look for environmental effects on galaxies and has a hint for the evolution of the most evolved galaxies. The morphology-density relation clearly shows the environmental dependence on galaxy morphology (Dressler 1980; Postman \& Geller 1984), which causes the expectation of finding more early-type galaxies in the massive clusters. Indeed, cluster redshift surveys have found an increasing early-type fraction with cluster velocity dispersion at high redshift (Postman 2005; Desai et al. 2007; Poggianti et al. 2009). In the local universe, however, the morphological dependence on the cluster mass is unclear. Simard et al. (2009) could not detect any trend between cluster velocity dispersion and early-type fraction based on the clusters from Sloan Digital Sky Survey (SDSS; York et al. 2000; Abazajian et al. 2009). Hoyle et al. (2012) also reported nearly constant early-type fraction against cluster halo mass. Galaxy morphology is an important product of both secular and non-secular evolution of galaxies, and so it contains a clue to the process that a galaxy has gone through. Investigating morphological contents of clusters allows us to understand the evolution of galaxies in the dense environment.

Rich clusters were found to be dominated by passive populations which are more peaked toward the center of clusters than star-forming galaxies (Biviano et al. 1997). Also, H$\rm I$ observations have reported that cluster galaxies are deficient in gas (i.e., Haynes, Giovanelli, \& Chincarini 1984; Boselli \& Gavazzi 2006; Jaffe et al. 2015). On the other hand, one can observe plenty of star-forming populations in clusters at high redshift (Butcher \& Oemler 1984; Couch \& Sharples 1987; Ellingson et al. 2001; Poggianti et al. 2006). The global trends of star-forming activities in low-z and high-z clusters suggest that galaxy clustering has played a significant role in quenching star formation and transforming galaxy morphology. However, how and when galaxies became passive in the cluster environment is still in question. As is well known from numerical simulations, galaxy-galaxy mergers are one of the most efficient ways for transforming galaxy morphology (e.g., Toomre \& Toomre 1972; Naab, Jesseit, \& Burket 2006). Galaxies can also lose their gas and truncate star formation by interacting with the intergalactic medium (ram-pressure stripping, e.g., Gunn \& Gott 1972; Smith et al. 2016) or the cluster potential (galaxy harassment, Moore, Lake, \& Katz 1998).

 Following the contemporary hierarchical formation scenario, massive galaxies have been assembled through major mergers as well as a number of minor mergers which continued until the recent epoch. The theoretical expectation has repeatedly been supported by inspecting field elliptical galaxies in deep images which show signatures indicating substantial recent merger activities (e.g., van Dokkum 2005; Kaviraj 2010). Galaxy clusters, however, are not likely places to see frequent mergers because of high peculiar motion of galaxies. Then, do mergers matter on the evolution of present-day cluster galaxies? If galaxies in clusters are free from the influence of galaxy-galaxy mergers, the evolution of cluster galaxies would be more predictable than that of field galaxies. Therefore, the frequency of galaxy mergers in the cluster environment is exceedingly important to predict the fate of cluster galaxies. There were some efforts to ascertain merger-related features in the cluster environment through deep optical images. Sheen et al. (2012) found post-merger features in 24\% of red-sequence galaxies based on four rich clusters; whereas Adams et al. (2012) report only 3\% of tidal features based on 54 galaxy clusters. These contradicting reports make it unclear whether mergers are responsible for the evolution of cluster galaxies even in the recent epoch. 
 
In this study, we aim to present the demographics of cluster galaxies regarding morphological contents, stellar populations, and signatures of mergers. We carried out the KASI-Yonsei Deep Imaging Survey of Clusters (KYDISC) which targets 14 Abell clusters at $0.015 \lesssim z \lesssim 0.144$ with the goal of detecting low surface brightness features ($\mu_{\rmag} \sim 27\, \sur $) in the cluster environment. Deep observations have been performed using the Inamori Magellan Areal Camera and Spectrograph (IMACS; Dressler et al. 2006) on the 6.5-meter Magellan Baade telescope and the MegaCam on the 3.6-meter Canada-France-Hawaii Telescope (CFHT). In Section 2, we present the details of observations and data reductions of the KYDISC. Cluster membership was determined based on redshift information from archival data and the multi-object spectroscopy using three instruments as outlined in Section 3. The catalog contains photometry, redshift, visual morphology, structural parameters from ellipse fit and a bulge/disk decomposition which is described in Section 4, and local density of 1409 cluster galaxies. In Section 5, we show and discuss morphologies, colors, and mergers of cluster galaxies based on the KYDISC catalog; in Section 6, we present a summary and our conclusions. In Appendix~\ref{app:cat}, we present the form and content of the KYDISC catalog.

Throughout the paper, we adopt Planck 2015 results with $\Omega_{\rm m} = 0.308$, $\Omega_\Lambda= 0.692$, and $H_0 = 67.8$ km/s/Mpc (Ade et al. 2016).

\section{Observations and data reduction}
\label{sec:data}
  \subsection{The deep optical imaging}
      \subsubsection{The Magellan Baade IMACS observation}
  The optical images of three Abell clusters (Abell 1146, Abell 3574, and Abell 3659) were obtained in $\gmag$ and $\rmag$ bands using the IMACS on the 6.5-meter Magellan Baade telescope at Las Campanas Observatory (LCO) on a dark night in 2012. We used a f/2 camera on the IMACS with Charge-Coupled Devices (CCDs) consisting of eight mosaic chips covering a $27^{\prime}$ circular area in diameter. Descriptions of target clusters and details of observation are shown in Table~\ref{tab:obs}. Four nearby fields of Abell 3574, the closest cluster in our data, were targeted and covered one fourth of the area including 0.85 virial radii of the cluster.
   
 All clusters were observed with two different exposures, long and short, for detection of low surface brightness features and a reliable photometry without light saturation, respectively. For long exposures, we targeted five-dithered points to recover the gaps in the mosaic and reduce the instrumental artifacts such as scattered light in the course of data reduction. The dithering is primarily necessary for the detection of low surface brightness features because faint features can be exposed only in the images where a variation in the background is well removed. Integrated exposure times are 1250 seconds for Abell 3574 and 2500 seconds for the others, respectively. The center of bright galaxies is saturated in the images with long exposure time, and therefore we took an additional image for each cluster with short exposure for an accurate estimation of magnitudes. 
 
 Calibration frames including bias and dome flats were taken during the daytime. Ten sky flats were acquired at twilight and dawn on the same day with appropriate exposure time according to sky flux at the moment. Five standard fields in the southern standard catalog\footnote{Southern Standard Stars for the $\umag , \gmag , \rmag , \imag , \zmag$ System: http://www-star.fnal.gov} (Smith et al. 2005) were also observed on the night for each filter in various airmass.

 \subsubsection{The CFHT MegaCam observation}
  Deep images of 11 Abell clusters in the $\umag$, $\gmag$, and $\rmag$ bands were obtained by queued service observations using the MegaCam on the 3.6-meter CFHT in the 2012A and 2013A seasons. The MegaCam with 36 CCDs has a large field of view (FOV) of 1 degree and covers one to three virial radii of the target clusters. The queued service observations have been performed by experts on the telescope following our requests on each exposure (e.g., exposure times, airmass, and sky conditions) and the constraints regarding image qualities. Specifically, our deep images were taken on dark nights, and the seeing varied between 0.65 to 0.8 arcseconds. We summarize the observation with the CFHT in Table~\ref{tab:obs}. These observations were carried out with the LSB mode on the CFHT. The most distinct aspect of the LSB mode is the large dithered pattern which enables us to have images with extremely flat background and detect extended light in low surface brightness. For the CFHT observation, we did not take additional short exposures, because our long exposure images with an exposure time of 420 seconds are safe from light saturation. The total integration time for each band was 3320 seconds for the $\umag$ band and 2940 seconds for the $\gmag$ and $\rmag$ bands, respectively. 
  
        \begin{table*}[ht]
        \centering
        \caption[Summary of deep observation]{Summary of deep photometry}
        \begin{tabular}{c l c c c c c c c}
        \hline \hline
        \multicolumn{1}{c}{Cluster ID} && RA(J2000) & Dec(J2000) & Instrument & Filter & {$t_{\rm{exp}}$}\tablenotemark{a}  & N of dithering & Date \\
                                                        &&hh:mm:ss    & dd:mm:ss     &     &       & sec          &             &    \\
        \hline
        Abell 1146 &      & 11:01:29.9 & -22:46:19 & IMACS & $\gmag$ & 2500 (100) & 5 & 2012/05/16\\
        ($z\sim$ 0.142)     & &       &                   &           & $\rmag$  & 2500 (100) &   5 \\ 
        Abell 3574 & -1 & 13:48:54.2  & -30:23:01 &           & $\gmag$ & 1250 (30)   &5    \\
        ($z\sim$ 0.015)    & &      &                   &           & $\rmag$  & 1250 (30)   &   5\\
            & -2 & 13:47:31.7 & -30:26:09 &           & $\gmag$  & 1250 (30)   &   5 \\
            &      &                    &                   &           & $\rmag$  & 1250 (30)   &   5 \\
            & -3 & 13:48:44.4 &  -30:43:59 &           & $\gmag$ & 1250 (30)   &    5\\
            &      &                    &                   &           & $\rmag$  & 1250 (30)   &    5\\
            & -4 & 13:47:19.3  & -30:48:42 &           & $\gmag$  & 1250 (30)  &  5  \\
            &      &                    &                   &           & $\rmag$  & 1250 (30)   &   5 \\            
         Abell 3659 &      & 20:02:27.4 & -30:04:50 &           & $\gmag$ & 2500 (100)  &5    \\
        ($z\sim$ 0.090)&      &                    &                   &           & $\rmag$  & 2500 (100)  & 5   \\ 
        \hline
        Abell 116   && 00:55:49.05 & 00:45:01.1  & MegaCam & $\umag$ & 3320 & 5 & 2012/07/26\\
        ($z\sim$ 0.067)& &                    &              &       & $\gmag$ & 2940 & 7 & 2012/08/21\\ 
                         &       &             &                     && $\rmag$ & 2940 & 7 & 2012/08/16\\         
        Abell 646   & &08:22:09.53 & 47:05:40.1  && $\umag$ & 3320 & 5 & 2013/01/14\\
        ($z\sim$ 0.127) &&                    &           &          & $\gmag$ & 2940 & 7 & 2013/01/14\\ 
                         &        &            &                     && $\rmag$ & 2940 & 7 & 2013/01/14\\                          
        Abell 655   & &08:24:50.00 & 46:51:40.1  && $\umag$ & 3320 & 5 & 2013/01/14\\
        ($z\sim$ 0.127) &&                    &           &          & $\gmag$ & 2940 & 7 & 2013/01/14\\ 
                         &        &            &                     && $\rmag$ & 2940 & 7 & 2013/01/14\\ 
        Abell 667   && 08:28:10.00 & 44:48:38.2  && $\umag$ & 3320 & 5 & 2012/10/20\\
        ($z\sim$ 0.144) &&                     &             &        & $\gmag$ & 2940 & 7 & 2012/10/21\\ 
                         &&                     &                     && $\rmag$ & 2940 & 7 & 2013/02/02\\ 
        Abell 690   & &08:39:16.12 & 28:57:02.7  && $\umag$ & 3250 & 5 & 2013/02/08\\
        ($z\sim$ 0.080) &&                    &              &       & $\gmag$ & 2940 & 7 & 2013/02/08\\ 
                         &        &            &                     && $\rmag$ & 2940 & 7 & 2013/02/13\\ 
        Abell 1126 && 10:53:50.90 & 16:57:01.7  && $\umag$ & 3320 & 5 & 2013/01/17\\
        ($z\sim$ 0.084) &&                    &                &     & $\gmag$ & 2940 & 7 & 2013/01/14\\ 
                         &   &                 &                     && $\rmag$ & 2940 & 7 & 2013/01/16\\ 
        Abell 1139 & &10:59:08.67 & 01:42:26.7  && $\gmag$ & 2940 & 7 & 2013/04/17\\
        ($z\sim$ 0.040) &&                    &              &       & $\rmag$ & 2940 & 7 & 2013/04/17\\ 
        Abell 1278 & &11:30:31.60 & 20:41:40.9  && $\umag$ & 3320 & 5 & 2013/02/14\\
        ($z\sim$ 0.134) & & &                   &                     & $\gmag$ & 2940 & 7 & 2013/04/09\\ 
                         &        &            &                     && $\rmag$ & 2940 & 7 & 2013/04/10\\ 
        Abell 2061 & &15:21:18.87 & 30:36:15.5 && $\umag$ & 3250 & 5 & 2013/05/12\\
        ($z\sim$ 0.078) &&                    &             &        & $\gmag$ & 2940 & 7 & 2013/05/12\\ 
                         &         &           &                     && $\rmag$ & 2940 & 7 & 2013/05/13\\ 
        Abell 2249 && 17:09:46.67 & 34:32:53.2 & & $\umag$ & 3250 & 5 & 2013/05/13\\
        ($z\sim$ 0.085) &&                    &             &        & $\gmag$ & 2940 & 7 & 2013/05/14\\ 
                         &          &          &                     && $\rmag$ & 2940 & 7 & 2013/06/05\\         
        Abell 2589 & & 23:24:01.50 & 16:56:23.0  &  & $\umag$ & 3320 & 5 & 2012/07/25\\
        ($z\sim$ 0.041) &&                    &                    & & $\gmag$ & 2940 & 7 & 2012/07/25\\ 
                                     &&                    &                    & & $\rmag$ & 2940 & 7 & 2012/07/26\\ 
        \hline \hline
        \end{tabular}
        \tablenotetext{1}{The number in parentheses refers to the exposure time of short-exposure images for the Magellan data.}
        \label{tab:obs}
      \end{table*}

 \subsection{Data Reduction}   
  \subsubsection{Preprocessing}
  The photometric data from the Magellan telescope were reduced using the Interactive Data Reduction and Analysis Facility (IRAF\footnote{IRAF is distributed by NOAO which is operated by AURA Inc., under cooperative agreement with NSF}) procedures for overscan correction, trimming, bias subtraction, and flat fielding based on ten skyflat images taken during twilight and dawn. 

  For the CFHT data, preprocessed images utilizing an Elixir pipeline were provided by CFHT team (Magnier \& Cuillandre 2004). The Elixir pipeline includes overscan correction, bias subtraction, and flat-fielding based on the large database of calibration frames. It also derives a photometric zero-point using standard stars.
 
 \subsubsection{Super sky subtraction}
 To detect low surface brightness features, we need a highly flat background. Even after applying preprocessing, we can observe small gradients and patterns on background perhaps due to the variations in the sky and detector response of each CCD. In the CFHT images, the effect of scattered light from the optics is prominent; this is known as bow structure causing 10\% of the sky background\footnote{http://www.cfht.hawaii.edu/Instruments/Elixir/scattered.html}. As a result, central chips are brighter than those on edge. Therefore, additional treatments for sky background are required for a more perfectly smooth background. 
 
  We generated the super sky frame for each cluster by combining all dithered frames after scaling background and masking sources brighter than 1$\sigma$ from the sky background, then subtracted this super sky frames from object frames.
   
  \subsubsection{Image co-addition}
We have updated the world coordinate system (WCS) of the Magellan data based on stars in USNO-B catalog (Monet et al. 2003), having low proper motions and similar magnitudes of our target galaxies. Five dithered images were co-added into a single deep image using the tasks in the MSCRED package in IRAF. 

 For the CFHT data, we calculated astrometric and photometric solutions using Software for Calibrating AstroMetry and Photometry (SCAMP; Bertin 2006). The photometric catalogs from Source Extractor (SExtractor; Bertin \& Arnouts 1996) and SDSS were used as input and reference catalogs, respectively. The seven dithered images with 36 CCDs each were combined using SWarp (Bertin et al. 2002). Images were resampled to have the fixed pixel scale of 0$''$.185/pixel using the distortion map from SCAMP. SWarp co-adds multi-extension images with the WCS information and flux scales between each dithered image from SCAMP. Finally, we obtained a deep image for each cluster and band.  The depth of the $\rmag$ band images is estimated to be 27 and 26.7 $\,\sur$ based on 3$\sigma$ of sky rms in 1 arcsec$^2$ bin for the Magellan and CFHT, respectively.
  
\begin{figure}
       \centering
       \includegraphics[width=0.5\textwidth]{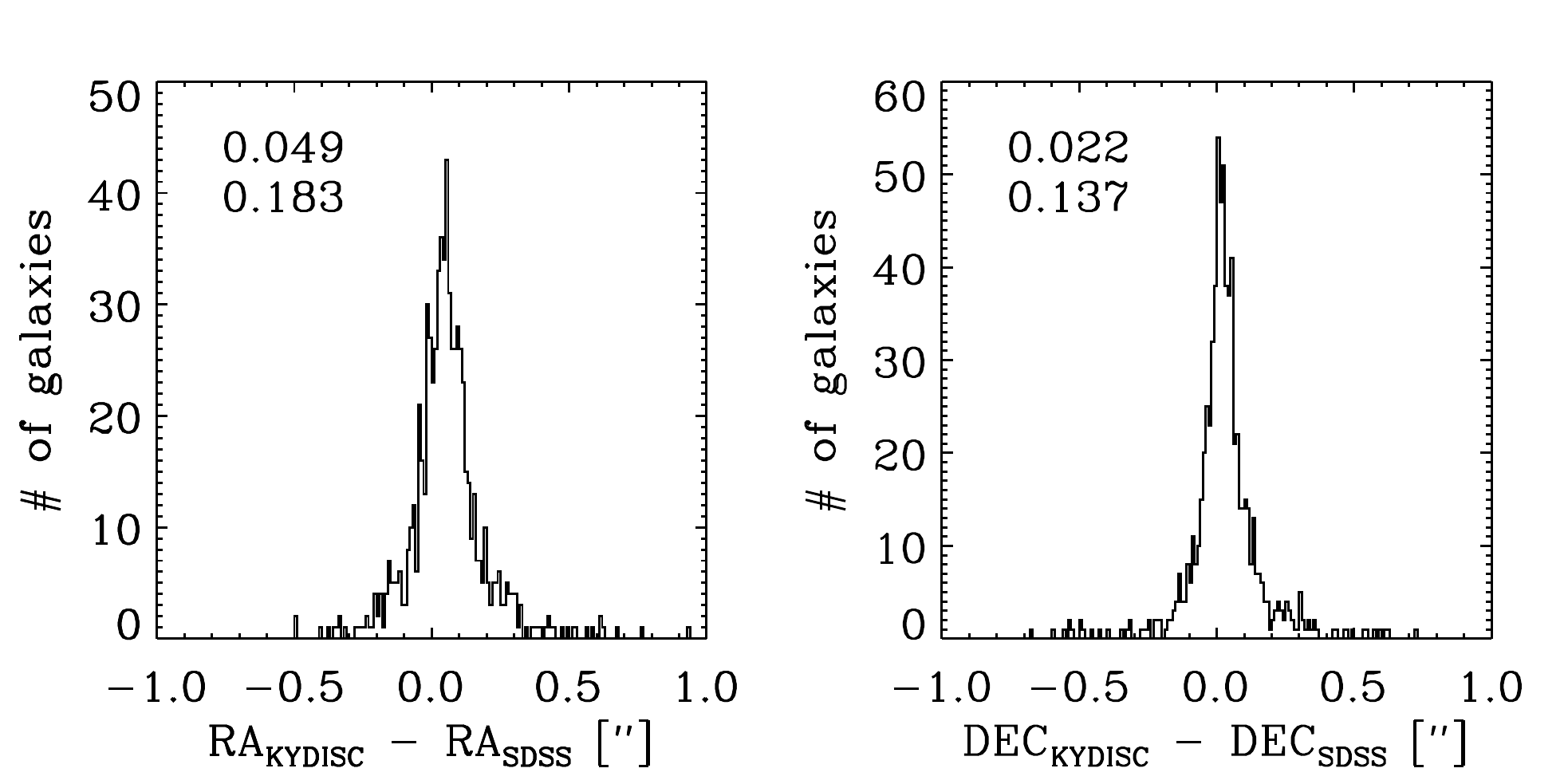}
       \caption[Position difference between KYDISC and SDSS.]
      {The differences in $RA$ and $DEC$ of galaxies between the KYDISC and SDSS. The median (top) and standard deviation (bottom) of the difference are shown in each panel.}
       \label{radiff}
     \end{figure}

Figure~\ref{radiff} shows the difference in Right Accession ($RA$) and Declination ($DEC$) between KYDISC and SDSS. The mean difference and standard deviation are approximately equal to one fourth and one pixel size (0$''$.185), respectively.

\subsection{Photometry}
\label{sec:phot}
  In our study, we adopted the auto magnitude system from SExtractor. For the Magellan data, we measured an instrumental magnitude mainly from short-exposure images to avoid light saturations in long-exposure images. Additionally, we added some galaxies which are located in CCD gaps into our sample from long-exposure images. We confirmed that target galaxies in the clusters observed using CFHT were free from the light saturations, and therefore we measured their instrumental magnitudes from co-added images. The $\umag$- and $\gmag$-band images were remapped to have the same WCS information with the $\rmag$-band image, which allows detecting objects in the $\rmag$ band and then measuring values in the $\umag$ and $\gmag$ bands. That is, the flux of each source was measured with the same aperture in all bands. 
 
We used the Southern standard stars observed on the night for standardization of the Magellan data. We selected 106 standard stars from the five standard fields which have a small magnitude error less than 0.02 magnitude both in the catalog and the measurement. The magnitudes of standard stars were measured using SExtractor with a $14^{\prime\prime}$ circular aperture which is the same aperture with the reference catalog from Smith et al. (2005). 
Standard magnitudes were acquired from instrumental magnitudes using the following steps. First, we estimated the airmass term from the slope of the least-squares fit of the magnitude difference between the catalog and the instrument against airmass. Then, we calculated the color term from the least-squares fit to the magnitude difference after applying the airmass term to the standard equation. The zero points are median values of the magnitude difference between the catalog and this observation after applying the airmass and color terms to the standardization equation. The standard deviation of the magnitude difference between the catalog and this observation is around 0.02 magnitude. 

\begin{figure}
       \centering
       \includegraphics[width=0.5\textwidth]{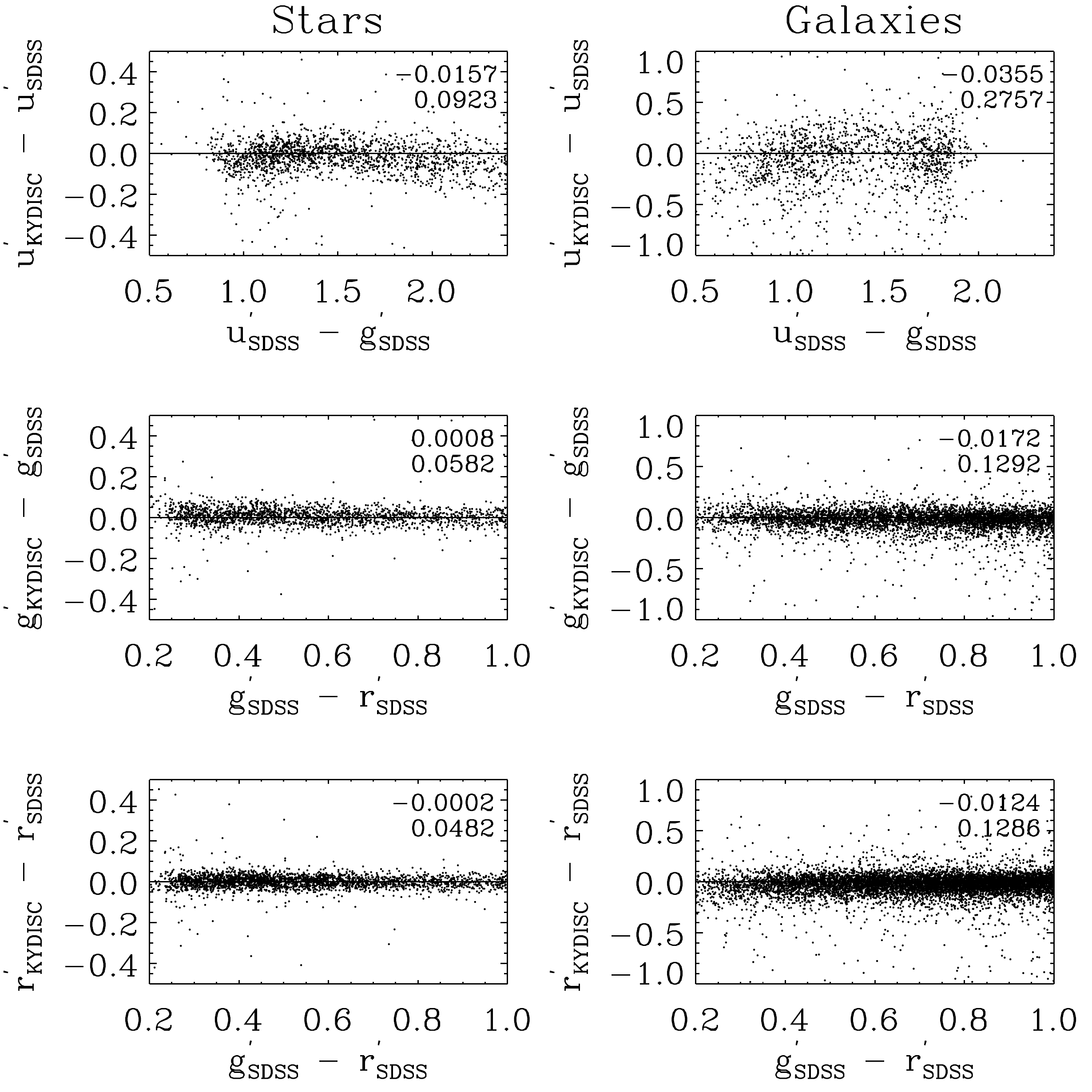}
       \caption[Magnitude difference between KYDISC and SDSS.]
      {Magnitude differences between KYDISC and SDSS as a function of SDSS colors. We used the SDSS PSF and SDSS Petrosian magnitudes for stars and galaxies, respectively. The median (top) and standard deviation (bottom) of the magnitude difference are shown in each panel. }
       \label{magdiff}
     \end{figure}

We adopted standardization coefficients from the Elixir pipeline by the CFHT team for the CFHT magnitudes. 
We estimated the photometric accuracy by comparing the KYDISC magnitudes with those from the SDSS (Figure~\ref{magdiff}). Stars and galaxies brighter than 20th magnitude in $\umag$, $\gmag$, and $\rmag$ bands were utilized for this comparison. The average differences between the KYDISC and the corresponding SDSS PSF magnitudes for stars are -0.0157, 0.0008, and 0.0002 in $\umag$, $\gmag$, and $\rmag$ bands, respectively. For galaxies, the average differences between the KYDISC and SDSS Petrosian magnitudes are -0.036, -0.017, and -0.012 in $\umag$, $\gmag$, and $\rmag$ bands, respectively. The magnitude differences between KYDISC and SDSS are small, and we also could not find a dependency of magnitude differences on colors. We conclude that KYDISC magnitude system reasonably agrees with the SDSS magnitude system.

Galactic foreground extinction was corrected using the dust map from Schlafly \& Finkbeiner (2011) assuming a Galactic extinction law with $A_{\rm V} / E(B-V)$ = 3.1 (Fitzpatrick 1999). We conducted k-correction following the algorithm described in Chilingarian, Melchior, \& Zolotukhin (2010) and Chilingarian \& Zolotukhin (2012) on magnitudes and colors.
      
 \subsection{Follow-up spectroscopy}
  \subsubsection{The Magellan IMACS}
   Galaxy spectra for Abell 1146 and Abell 3659 were taken using the IMACS in 2013. We used the 200 lines/mm grism with GG 495 filter which covers 4800--7800 $\AA$ with a resolution of R $\sim$ 700.
   We allocated two slit masks for each cluster targeting galaxies brighter than 20th magnitude in the $\rmag$ band and obtained 269 and 246 galaxy spectra in Abell 1146 and Abell 3659, respectively. Each mask contains over 100 slits with a width of 1$''$ and a length of 12$''$. The exposure times were $3 \times 1800$ seconds for each mask except the second mask of Abell 3659 whose exposure time was $3 \times 1300$ seconds. Each time we observed target spectra, we took an arc and three dome flat frames for calibrations. The multi-object spectroscopy (MOS) observations are summarized in Table~\ref{tab:spec}.
  
  The Carnegie Observatories System for MultiObject Spectroscopy (COSMOS\footnote{http://code.obs.carnegiescience.edu/cosmos}) was utilized for the reduction of the IMACS MOS data. It automatically finds each spectrum using observation definition files and performs wavelength calibration with known spectral lines of HeNeAr lights. We subtracted sky spectra from object frames after performing bias subtraction and flat-fielding. Galaxy spectra were extracted and combined into a single spectrum.

      \begin{table}
        \centering
        \caption[Summary of the follow-up spectroscopy]{Summary of the follow-up spectroscopy}
        \begin{tabular}{c c c c c c c c c c c c}
        \hline \hline
        \multicolumn{1}{c}{Cluster ID} & Mask & Instrument & $t_{\rm{exp}}$  & N of target \\
                                                        &   &           & sec          &             \\
        \hline
        Abell 1146 &    1  &IMACS& 3$\times$1800 & 147 \\
                         &    2  && 3$\times$1800 & 122 \\
         Abell 3659 &     1 &&  3$\times$1800  & 131 \\
                         &    2  && 3$\times$1300 & 115 \\
       \hline
        Abell 3574 & 1 & WFCCD & 2$\times$1500 & 27 \\
                         & 2 &  & 2$\times$1500 & 29 \\
                         & 3 &  & 2$\times$1600 & 28 \\
                         & 4 &  & 2$\times$1600 & 30 \\
                         & 5 &  & 2$\times$1600 & 37 \\
\hline
        Abell 1126 & 1 & Hydra & 2$\times$1200 & 71 \\
                          & 2 &           & 2$\times$1500 & 70 \\
                          & 3 &           & 2$\times$1300 & 69 \\
                          & 4 &           & 3$\times$1200 & 64 \\
                          & 5 &           & 2$\times$1800 & 68 \\
        Abell 2061 & 1 &          & 2$\times$1500 & 73 \\
                          & 2 &           & 2$\times$1500 & 70 \\
                          & 3 &           & 2$\times$1800 & 51 \\
                          & 4 &           & 2$\times$1800 & 60 \\
        Abell 2249 & 1 &          & 2$\times$1600 & 65 \\
                         & 2 &           & 3$\times$1400 & 63 \\
                         & 3 &           & 2$\times$1800 & 67 \\
                         & 4 &           & 3$\times$1200 & 57 \\
                         & 5 &           & 3$\times$1200 & 54 \\
        \hline \hline
        \end{tabular}
        \label{tab:spec}
      \end{table}

  \subsubsection{du Pont WFCCD}
   We carried out multi-slit spectroscopy for Abell 3574 with the Wide-Field Reimaging CCD Camera (WFCCD) on the du Pont 2.5-meter telescope at LCO in 2014. The WFCCD with the blue grism covers 3800--7600 $\AA$ with a resolution of R $\sim$ 970 and is capable of MOS over 25$'$ $\times$ 25$'$ FOV, simultaneously. Abell 3574 fields were covered by the 6dF survey; however, only 33 bright galaxies were observed. We targeted 124 galaxies brighter than 19th magnitude in the $\rmag$ band. Five multi-slit masks including slits with a width of 1$''$.5 and a length greater than 12$''$ were generated. We took 20 bias frames during the daytime. An arc frame with a HeNe lamp and three dome flat frames were taken before or after each object exposure. The integrated exposure times varied from 2400 to 4200 seconds depending on the target magnitude of each mask and the weather condition. We additionally observed a radial velocity standard using the same width slit. 
  
  The multi-slit data were reduced mainly using the modified version of the WFCCD reduction package (Prochaska et al. 2006; Lim et al. 2015). First, we applied overscan correction, bias subtraction, and flat fielding to the raw images. Slit spectra are distorted as they pass through optics. These curvatures in spectra were calculated from flat frames, and each spectrum was extracted with the curvature information. Then, the wavelength calibration is done with the arc frames. Cosmic rays were removed using $LA\,Cosmic$ (van Dokkum 2001) which detects cosmic rays and replaces pixel values with interpolation of neighboring pixels. Finally, we subtracted sky spectra from each galaxy, and extracted 1-D spectra of target galaxies.
    
  \subsubsection{WIYN Hydra}
  We obtained galaxy spectra for Abell 1126, Abell 2061, and Abell 2249 using the Hydra on the WIYN\footnote{The WIYN Observatory is a joint facility of the University of Wisconsin-Madison, Indiana University, the National Optical Astronomy Observatory and the University of Missouri.} 3.5-meter telescope in 2015. We used 600 lines/mm grating with red cable which consists of 90 2$''$-diameter fibers. A spectral window is 4579--7431 $\AA$ with a resolution of R $\sim$ 1790. The large FOV of the Hydra (1$\deg$), which is the same as the MegaCam, enables us to obtain many spectra for these clusters efficiently. We set 4-5 fiber configurations for each cluster targeting bright galaxies ($M_{\rmag}  < -19$) without redshift information on archival data. Ten to twelve fibers in each configuration were assigned to the position of a blank sky to take sky spectra. Each fiber configuration was taken two or three times according to the overall magnitude of galaxies in each configuration and the weather condition (Table~\ref{tab:spec}). An arc frame and three dome flats were also obtained for each fiber configuration. 

The raw spectra from the Hydra were reduced for preprocessing, wavelength calibration, and sky subtraction employing the $dohydra$ routine in IRAF. 

 \subsubsection{Radial velocity}
 \label{sec:red}
Galaxy spectra were resampled to have a pixel size of 2$\AA$. Galaxies having an S/N per $\AA$ lower than 5 were excluded from the sample to only consider reliable measurements, which eliminated five galaxies from the cluster sample described in Section 3. We measured the radial velocity of each galaxy by using the $fxcor$ task in IRAF which returns the best-matched radial velocities employing the Fourier cross-correlation method described in Tonry \& Davis (1979). The spectrum of the radial velocity standard star was used as the template of known redshift. The wavelength ranges containing strong sky absorptions were excluded in this analysis: 3800--5500; 5600--6200; 6400--6800; and 7000--7500 $\AA$. The typical errors on radial velocity were 33, 45, and 38 km/s for the IMACS, WFCCD, and Hydra data, respectively.

\section{Membership}
We determined cluster memberships based on spectroscopic information from the observations using the IMACS, Hydra, and WFCCD, and from archival data. We collected redshift information from the NASA/IPAC Extragalactic Database (NED) and the SDSS for the sample cluster fields. Additionally, we obtained redshift information for Abell 667, Abell 655, and Abell 646 from Hectospec Cluster Survey (HeCS; Rines et al. 2013).

     \begin{figure*}
       \centering
       \includegraphics[width=1\textwidth]{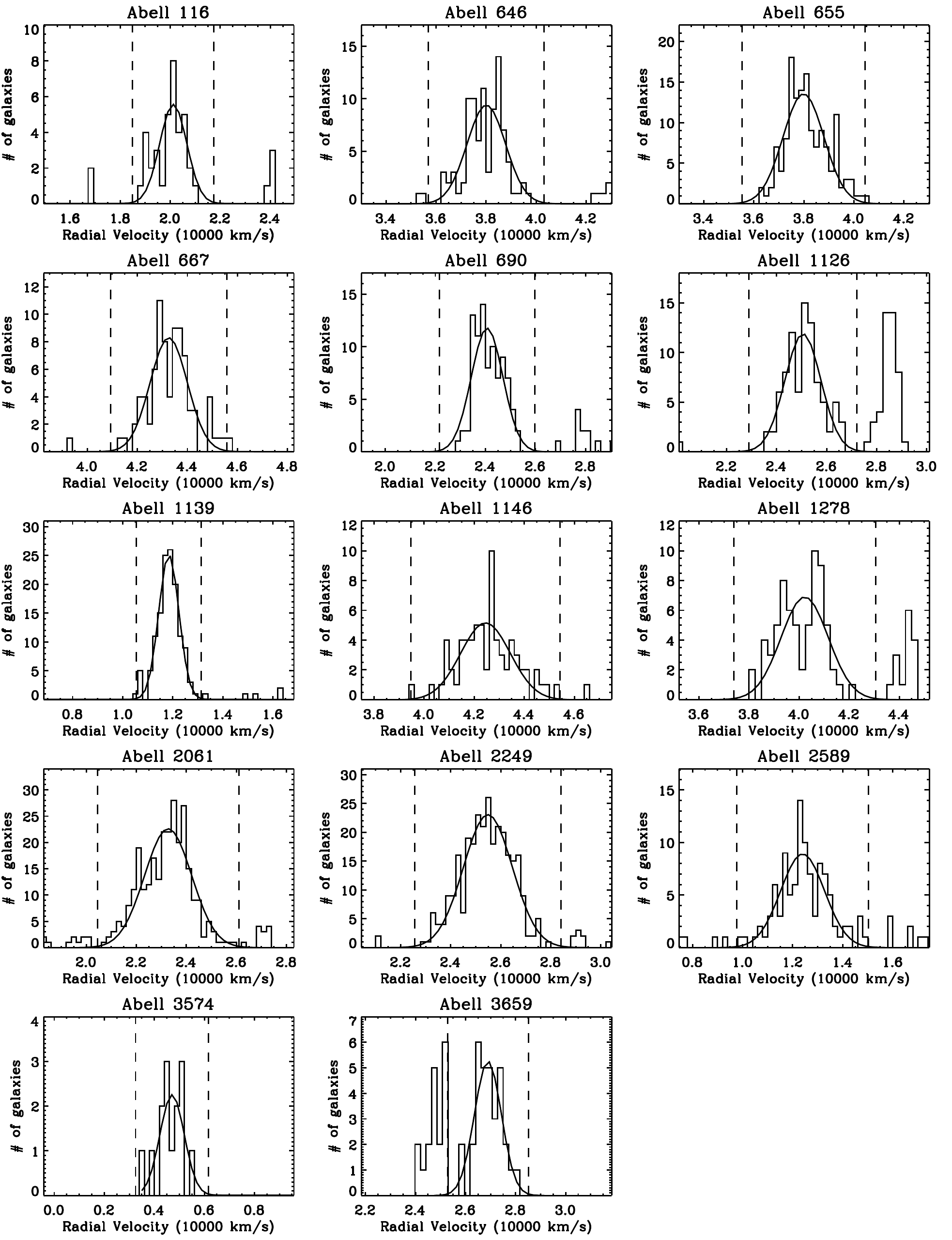}
       \caption[The radial velocity distribution]
      {The radial velocity distribution for each cluster. The dashed lines indicate $v_{\rm cen}\pm\,3\sigma_{\rm obs}$ which is one of the membership criteria in this study.}
       \label{rv1}
     \end{figure*}

The radial velocity dispersion ($\sigma_{\rm obs}$) and the central radial velocity ($v_{\rm cen}$) of clusters are estimated from a Gaussian fit of the distribution of radial velocities (Figure~\ref{rv1}). The redshift of each cluster ($z_{\rm cl}$) is calculated from $v_{\rm cen}$, $z_{\rm cl} = {v_{\rm cen}/ c}$, where $c$ is the speed of the light. We selected galaxies as possible members of each cluster when they have a radial velocity ($v_{\rm rv}$) within $\pm 3\,\sigma_{\rm obs}$ from the central radial velocity ($v_{\rm cen}$).

The virial radius ($R_{200}$) defining a sphere with a mean density of 200 times critical density is estimated using the formula from Carlberg, Yee, \& Ellingson (1997):
  \begin{equation}
     R_{200} = {\sqrt{3} \over 10} {\sigma_{\rm cl} \over H(z) } = {\sqrt{3} \over 10} {\sigma_{\rm cl} \over {H_0 \sqrt{\Omega_m (1+z_{\rm cl})^3 + \Omega_\Lambda}} }
   \end{equation}
where $\sigma_{\rm cl}$ is the radial velocity dispersion corrected for the cluster redshift, $\sigma_{\rm cl} = \sigma_{\rm obs} / (1 + z_{\rm cl})$. 
The corresponding cluster mass ($M_{200}$) was estimated as follows:
    \begin{equation}
      M_{200} ={ 4\over 3 }\pi \,200\, \overline{\rho} \,{R_{200}}^3 = 100 \,H_0^2 \,\Omega_{\rm m}\, R_{200}^3 \,/\, G 
    \end{equation}
where $G$ is the gravitational constant.
The cluster-centric distance ($R$) of each galaxy from the brightest cluster galaxy (BCG) was measured, and we identified galaxies that locate within two times $R_{200}$. In summary, we selected galaxies with $|v_{\rm rv} - v_{\rm cen}| < 3\,\sigma_{\rm obs}$,  $R<2R_{200}$, and $M_{\rmag} < -19.8$ as the member of each cluster. We present characteristics of clusters derived from spectroscopic data in Table~\ref{tab:cltinfo}. 

The spectroscopic completeness for the galaxies brighter than $M_{\rmag}=-$19.8 derived by assuming that galaxies are at the cluster redshift is varied in clusters (Table~\ref{tab:comp}). We note that we have spectroscopic information only for galaxies brighter than $M_{\rmag}\sim -21.3$ in Abell 1278. The spectroscopic completeness of the survey is 0.8 excluding Abell 1278.
      
\begin{table}
  \centering
  \caption[Cluster characteristics]{Cluster characteristics}
  \begin{tabular}{ccccccc}
  \hline \hline
  \multicolumn{1}{c}{Cluster} & $v_{\rm cen}$ & $\sigma_{\rm obs}$ & $R_{200}$  & $M_{200}$ & {$N_m$}\tablenotemark{a}\\
   & km/s & km/s & Mpc & $10^{14} M_{\odot}$&  & \\
  \hline
  Abell 116 & 20123 & 542$\pm$68 & 1.26$\pm$0.17 & $2.25^{+0.79}_{-1.02}$ & 44 \\
  Abell 646 & 38007 & 771$\pm$56 & 1.64$\pm$0.14 & $5.33^{+1.21}_{-1.42}$ & 98 \\
  Abell 655 & 37985 & 821$\pm$44 & 1.75$\pm$0.11 & $6.43^{+1.11}_{-1.24}$ & 180 \\
  Abell 667 & 43255 & 775$\pm$59 & 1.61$\pm$0.14 & $5.12^{+1.23}_{-1.45}$ & 117 \\
  Abell 690 & 24068 & 633$\pm$43 & 1.44$\pm$0.11 & $3.42^{+0.71}_{-0.82}$ &  88 \\
  Abell 1126 & 25054 & 720$\pm$52 & 1.63$\pm$0.13 & $4.99^{+1.09}_{-1.27}$ &  73\\
  Abell 1139 & 11844 & 433$\pm$30 & 1.04$\pm$0.08 & $1.25^{+0.26}_{-0.29}$ &  52\\
  Abell 1146 & 42451 & 992$\pm$86 & 2.07$\pm$0.21 & $10.83^{+2.94}_{-3.58}$ &  121\\
  Abell 1278 & 40223 & 943$\pm$110 & 1.99$\pm$0.26 & $9.54^{+3.31}_{-4.30}$ &  41\\
  Abell 2061 & 23271 & 942$\pm$35 & 2.15$\pm$0.09 & $11.40^{+1.33}_{-1.44}$ &  250\\
  Abell 2249 & 25487 & 976$\pm$38 & 2.21$\pm$0.10 & $12.39^{+1.54}_{-1.66}$ &  240\\
  Abell 2589 & 12397 & 877$\pm$61 & 2.11$\pm$0.15 & $10.40^{+2.10}_{-2.42}$ &  67\\
  Abell 3574 & 4709 & 486$\pm$50 & 1.21$\pm$0.13 & $1.92^{+0.55}_{-0.66}$ &  14\\
  Abell 3659 & 26894 & 538$\pm$76 & 1.21$\pm$0.19 & $2.04^{+0.81}_{-1.09}$ &  24 \\
  \hline \hline
  \end{tabular}
  \label{tab:cltinfo} 
  \tablenotetext{1}{Spectroscopically-confirmed member galaxies.}
\end{table}

\begin{table}
 \centering
 \caption[Spectroscopic completeness]{Spectroscopic completeness}
 \begin{tabular}{ccccc}
 \hline \hline
 \multicolumn{1}{c}{Cluster} & {FOV [$R_{200}$]}\tablenotemark{a} & $N_{\rm phot}\tablenotemark{b}$ & {$N_{\rm spec}$}\tablenotemark{c} & Completeness\\
 \hline
   Abell 116 & 2.00 & 127 & 125 & 0.98 \\
   Abell 646 & 2.00 & 252 & 195 & 0.77 \\
   Abell 655 & 2.00 & 503 & 313 & 0.62 \\
   Abell 667 & 2.00 & 430 & 271 & 0.63 \\
   Abell 690 & 2.00 & 200 & 173 & 0.87 \\
   Abell 1126 & 2.00 & 275 & 251 & 0.91 \\
   Abell 1139 & 2.00 & 58 & 58 & 1.00 \\
   Abell 1146 & 1.35 & 191 & 151 & 0.79 \\
   Abell 1278\tablenotemark{d} & 2.00 & 809 & 327 & 0.40 \\
   Abell 2061 & 2.00 & 452 & 420 & 0.93 \\
   Abell 2249 & 2.00 & 471 & 414 & 0.88 \\
   Abell 2589 & 1.25 & 109 & 83 & 0.76 \\
   Abell 3574\tablenotemark{e} & 0.85 & 14 & 14 & 1.00 \\
   Abell 3659 & 1.70 & 56 & 47 & 0.84 \\
 \hline
  Total     && 3947 & 2842 & {0.72}\tablenotemark{f}\\
 \hline \hline
 \end{tabular}
  \tablenotetext{1}{Survey FOV for both imaging and spectroscopy.}
  \tablenotetext{2}{Galaxies within the FOV ($M_{\rmag}  < -19.8$).}
  \tablenotetext{3}{Galaxies with spectroscopic redshift ($M_{\rmag}  < -19.8$).}
  \tablenotetext{4}{Only bright galaxies ($M_{\rmag}< -21.3$) have spectroscopic redshifts.}
  \tablenotetext{5}{KYDISC covered one fourth of the area including 0.85$R_{200}$.}
  \tablenotetext{6}{Spectroscopic completeness is 0.8 excluding Abell 1278.}
  \label{tab:comp}
\end{table}

 \section{Analysis}
  \subsection{Fitting}
The $\rmag$ deep images were primarily used for both $ellipse$ fit and {\sc{galfit}}. We utilized a short-exposure image when we detect a light saturation of a target galaxy in a long-exposure image.
  
 \subsubsection{The ellipse fitting}
\label{sec:ell}
  Based on the $\rmag$-band images, we measured the effective radius ($r_{\rm e}$), position angle ($PA$), and ellipticity ($\epsilon = 1-b/a$) using the $ellipse$ task in IRAF STSDAS which models the light profile of a galaxy with elliptical isophotes. The center of the galaxies was fixed; however, we allowed free estimation of the position angle and ellipticity at a given level of light. All the contaminants were masked out based on a segmentation map from SExtractor to consider only the light from the target galaxy. We used the curve of growth of light from the radial intensity profile to measure the effective (half-light) radius ($r_{\rm e}$) and the radius containing 90\% of the total light ($r_{90}$). Both $r_{\rm e}$ and $r_{\rm 90}$ are corrected to be the circular one, $\overline{r_{\rm e}} = \sqrt{r_{\rm e}^2 \, (1-\epsilon_{\rm e}) }$ and $\overline{r_{\rm 90}} = \sqrt{r_{\rm 90}^2 \, (1-\epsilon_{\rm 90}) }$. We also constructed model images from the $ellipse$ fit using the $bmodel$ task in IRAF and subtracted them from the original images.

 \subsubsection{Two-dimensional decomposition using {\sc{galfit}}}
 \label{sec:gal}

 Galaxies were decomposed into bulge and disk components assuming a combination of free S\'{e}rsic and exponential (n = 1) profiles using the {\sc{galfit}} software (Peng et al. 2002, 2010). We built a point-spread function with bright but not saturated stars in each field. The same object masks that we have utilized in $ellipse$ fit were applied not to weight the light from other sources. Also, the structural informations from $ellipse$ fit were used as initial guesses (e.g., PA, $b/a$). 
 
 Mathematically-motivated fitting does not always produce physically meaningful solutions. Therefore, constraints allowing a specific range of each parameter are often used for decomposing galaxies into multiple components (e.g., MacArthur et al. 2003; Gadotti 2009; Weinzirl et al. 2009; Fisher \& Drory 2010; Meert et al. 2015; Gao \& Ho 2017). We set empirically-confirmed constraints for each parameter following Kim et al. (2016) to exclude physically meaningless solutions. The x and y centers of both bulge and disk components were fixed to be the same. The S\'{e}rsic index of a bulge component can vary from 1 to 8. Apparent axis ratio ($b/a$) of a bulge component is fitted within the range 0.3 to 1. We adopted a disk $b/a$ from the $ellipse$ fit with only allowing a small variation ($\pm\,0.1$). Position Angle of a disk component is fitted within the range $\pm\,20$ degrees from the initial guess.

 \subsection{Visual inspection}
 \label{sec:vis}
 
\begin{figure*}
       \centering
       \includegraphics[width=1\textwidth]{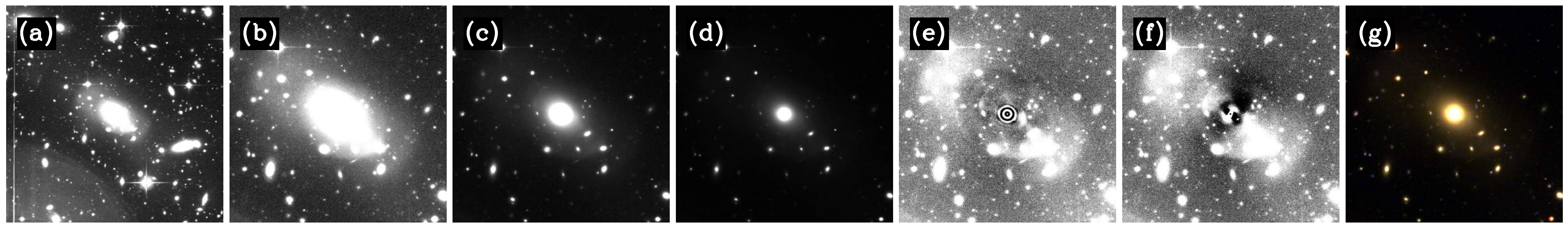}
       \caption[Sample image set for visual inspection]
      {Sample cutout image set for visual inspection. (a) $\rmag$-band cutout image covering eight times $\overline{r_{90}}$. (b)--(d) $\rmag$-band cutout images with different contrasts covering four times $\overline{r_{90}}$. (e) residual image from ellipse fitting. (f) residual image from {\sc {galfit}} model. (g) 2 ($\gmag$ and $\rmag$) or 3 ($\umag$, $\gmag$, and $\rmag$) color-composite image. More example cutout sets can be found in Figure~\ref{appvis} in Appendix~\ref{app:vis}.}
       \label{samvis}
     \end{figure*}

We visually classified galaxy morphologies based on the Hubble classification and merger features using a set of seven cutout images (Figure~\ref{samvis}). Figure~\ref{samvis} (a) is a large FOV cutout image. Figure~\ref{samvis} (b), (c), and (d) are $\rmag$-band cutout images with different contrasts which enabled us to give a focus on both outskirts and the inner region of a galaxy.  Figure~\ref{samvis} (e) is a residual image from $ellipse$ fit which often reveals asymmetric features. Figure~\ref{samvis} (f) is a residual image from {\sc {galfit}} which helps to detect lights beyond the physical models. Figure~\ref{samvis} (g) is a color composite image for detecting features with distinct colors from a galaxy body (e.g., collisional rings, dust lanes). We used $\gmag$, $\rmag$ color-composite image for four clusters (Abell 1139, Abell 1146, Abell 3574, and Abell 3659) and $\umag$, $\gmag$, $\rmag$ color-composite image for the other clusters.

Based on a set of seven cutout images, five researchers have independently performed a visual inspection without knowing any kinds of galaxy's information such as redshift, magnitude, size, cluster-centric distance, and B/T which can give some bias to judgment. Finally, the visual class of each galaxy is determined to be the most frequent decision by five researchers. Table~\ref{tab:sig} presents the significance of the classification ($N_{\rm S}$), the number of researchers who made the same decision with the final classification. Out of 1409 galaxies, 66 and 48 galaxies show the same frequency ($N_{\rm S}$ = 2) in two classes for the Hubble and merger classifications, respectively. In those cases, the final class was determined by following the decision of one researcher whose classification shows the highest agreement with the final classification: 84\% and 92\% for the Hubble and merger classifications, respectively.

 \begin{table}
\centering
\caption[Significance of the classification]{Significance of the classification}
\begin{tabular}{l c c c c c c} 
\hline \hline
	& Total & $N_{\rm S}= 5$& $N_{\rm S}=4$ & $N_{\rm S}=3$ & $N_{\rm S}=2$\\
	\hline
Hubble type & 1409 & 399 & 443 & 501 & 66 \\
Merger type & 1409 & 583 & 395 & 383 & 48 \\

 \hline
\hline
\end{tabular}
\label{tab:sig}
\end{table}

 \subsubsection{Morphological classification}
We divided galaxies into four groups based on the Hubble classification: ellipticals (E); lenticulars (S0); early spirals (S$_{\rm E}$); and late spirals (S$_{\rm L}$). We first identified early types (E/S0) showing smooth light profiles without spiral arms. Classifying E and S0 is complicated, and therefore careful and clear criterion is required to minimize subjectivity. When galaxies show only one component in any contrast images they were classified into E. Lenticular galaxies having both bulge and disk components sometimes reveal edge-on disk on residual images (Figure~\ref{samvis} (e) and (f)). S0 galaxies having face-on disk tend to show rapidly (bulge) and more mildly shrinking (disk) components through Figure~\ref{samvis} (b), (c), and (d) due to different light profiles of two components.
  
Late-type galaxies either having spiral arms or prominent disk components ($\geq40\%$) were divided into early (S$_{\rm E}$) and late (S$_{\rm L}$) spirals according to the dominance of two components. S$_{\rm E}$ types including Sa and Sb types in the standard Hubble classification have prominent bulge components which are brighter than disks and whose size is seemingly larger than 20\% of galaxy size in large contrast images (Figure~\ref{samvis} (c) and (d)). S$_{\rm L}$ types are disk-dominated galaxies including galaxies from Sc to Irr types whose bulge brightness is competitive with or fainter than disk brightness, and bulge size is smaller than 20\% of galaxy size especially in Figure~\ref{samvis} (c) and (d). We present sample images for our classification based on Hubble types in Figure~\ref{hcat}. Example cutout sets for each morphological type are shown in Figure~\ref{appvis} in Appendix~\ref{app:vis}.
 
  Galaxy clusters are the site where the most massive galaxies reside; therefore, we found a large fraction of early-type galaxies in our sample (Table~\ref{tab:mor}). The bulge fraction becomes larger from late to early types (Hudson et al. 2010); therefore, galaxy morphology has a close connection to the bulge-to-total ratio (B/T). We compared our classification with B/T in the $\rmag$ band from {\sc{galfit}} (Figure~\ref{morbt}). The mean values of the B/T were 0.78, 0.61, 0.33, and 0.15 for E, S0, S$_{\rm E}$, and S$_{\rm L}$ classes, respectively. Our result is comparable with Kim et al. (2016) who compared the B/T with visual classification from Khim et al. (2015). 

\begin{table}
\centering
\caption[Galaxy morphology based on visual inspection]{Galaxy morphology based on visual inspection}
\begin{tabular}{l c c c c} \hline \hline
Cluster	& Hubble type & &	\multicolumn{2}{c}{Merger type}\\
      &  & N  & OM &  PM\\	  
\hline
Abell 116 (44) & E (11) & 0.82 (9) & 0.00 (0) & 0.18 (2) \\
 & S0 (13) & 0.92 (12) & 0.00 (0) & 0.08 (1) \\
 & S$_{\rm E}$ (11) & 0.82 (9) & 0.00 (0) & 0.18 (2) \\
 & S$_{\rm L}$ (9) & 0.33 (3) & 0.11 (1) & 0.56 (5) \\
\hline
Abell 646 (98) & E (43) & 0.91 (39) & 0.00 (0) & 0.09 (4) \\
 & S0 (37) & 0.81 (30) & 0.00 (0) & 0.19 (7) \\
 & S$_{\rm E}$ (9) & 0.56 (5) & 0.00 (0) & 0.44 (4) \\
 & S$_{\rm L}$ (9) & 0.78 (7) & 0.00 (0) & 0.22 (2) \\
\hline
Abell 655 (180) & E (71) & 0.89 (63) & 0.00 (0) & 0.11 (8) \\
 & S0 (63) & 0.78 (49) & 0.00 (0) & 0.22 (14) \\
 & S$_{\rm E}$ (26) & 0.46 (12) & 0.04 (1) & 0.50 (13) \\
 & S$_{\rm L}$ (20) & 0.70 (14) & 0.00 (0) & 0.30 (6) \\
\hline
Abell 667 (117) & E (42) & 0.90 (38) & 0.05 (2) & 0.05 (2) \\
 & S0 (38) & 0.89 (34) & 0.00 (0) & 0.11 (4) \\
 & S$_{\rm E}$ (19) & 0.37 (7) & 0.00 (0) & 0.63 (12) \\
 & S$_{\rm L}$ (18) & 0.50 (9) & 0.06 (1) & 0.44 (8) \\
\hline
Abell 690 (88) & E (27) & 0.89 (24) & 0.00 (0) & 0.11 (3) \\
 & S0 (37) & 0.73 (27) & 0.03 (1) & 0.24 (9) \\
 & S$_{\rm E}$ (12) & 0.58 (7) & 0.00 (0) & 0.42 (5) \\
 & S$_{\rm L}$ (12) & 0.42 (5) & 0.08 (1) & 0.50 (6) \\
\hline
Abell 1126 (73) & E (22) & 0.86 (19) & 0.05 (1) & 0.09 (2) \\
 & S0 (31) & 0.65 (20) & 0.06 (2) & 0.29 (9) \\
 & S$_{\rm E}$ (15) & 0.60 (9) & 0.07 (1) & 0.33 (5) \\
 & S$_{\rm L}$ (5) & 0.60 (3) & 0.20 (1) & 0.20 (1) \\
\hline
Abell 1139 (52) & E (13) & 0.92 (12) & 0.08 (1) & 0.00 (0) \\
 & S0 (15) & 0.87 (13) & 0.00 (0) & 0.13 (2) \\
 & S$_{\rm E}$ (11) & 0.64 (7) & 0.27 (3) & 0.09 (1) \\
 & S$_{\rm L}$ (13) & 0.54 (7) & 0.08 (1) & 0.38 (5) \\
\hline
Abell 1146 (121) & E (66) & 0.83 (55) & 0.06 (4) & 0.11 (7) \\
 & S0 (37) & 0.73 (27) & 0.08 (3) & 0.19 (7) \\
 & S$_{\rm E}$ (17) & 0.53 (9) & 0.06 (1) & 0.41 (7) \\
 & S$_{\rm L}$ (1) & 1.00 (1) & 0.00 (0) & 0.00 (0) \\
\hline
Abell 1278 (41) & E (16) & 0.81 (13) & 0.06 (1) & 0.13 (2) \\
 & S0 (10) & 0.70 (7) & 0.00 (0) & 0.30 (3) \\
 & S$_{\rm E}$ (10) & 0.60 (6) & 0.00 (0) & 0.40 (4) \\
 & S$_{\rm L}$ (5) & 0.20 (1) & 0.40 (2) & 0.40 (2) \\
\hline
Abell 2061 (250) & E (79) & 0.90 (71) & 0.04 (3) & 0.06 (5) \\
 & S0 (92) & 0.82 (75) & 0.03 (3) & 0.15 (14) \\
 & S$_{\rm E}$ (58) & 0.50 (29) & 0.09 (5) & 0.41 (24) \\
 & S$_{\rm L}$ (21) & 0.67 (14) & 0.00 (0) & 0.33 (7) \\
\hline
Abell 2249 (240) & E (94) & 0.88 (83) & 0.03 (3) & 0.09 (8) \\
 & S0 (110) & 0.81 (89) & 0.01 (1) & 0.18 (20) \\
 & S$_{\rm E}$ (22) & 0.55 (12) & 0.18 (4) & 0.27 (6) \\
 & S$_{\rm L}$ (14) & 0.71 (10) & 0.00 (0) & 0.29 (4) \\
\hline
Abell 2589 (67) & E (15) & 0.93 (14) & 0.00 (0) & 0.07 (1) \\
 & S0 (41) & 0.85 (35) & 0.02 (1) & 0.12 (5) \\
 & S$_{\rm E}$ (6) & 0.67 (4) & 0.00 (0) & 0.33 (2) \\
 & S$_{\rm L}$ (5) & 0.60 (3) & 0.00 (0) & 0.40 (2) \\
\hline
Abell 3574 (14) & E (2) & 0.00 (0) & 0.50 (1) & 0.50 (1) \\
 & S0 (3) & 0.33 (1) & 0.33 (1) & 0.33 (1) \\
 & S$_{\rm E}$ (6) & 1.00 (6) & 0.00 (0) & 0.00 (0) \\
 & S$_{\rm L}$ (3) & 1.00 (3) & 0.00 (0) & 0.00 (0) \\
\hline
Abell 3659 (24) & E (5) & 0.80 (4) & 0.00 (0) & 0.20 (1) \\
 & S0 (10) & 0.70 (7) & 0.10 (1) & 0.20 (2) \\
 & S$_{\rm E}$ (9) & 0.56 (5) & 0.22 (2) & 0.22 (2) \\
 & S$_{\rm L}$ (0) & 0.00 (0) & 0.00 (0) & 0.00 (0) \\
\hline

\end{tabular}
\label{tab:mor}
\end{table}

       \begin{figure}
       \centering
       \includegraphics[width=0.45\textwidth]{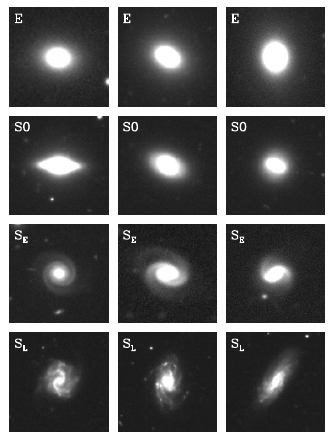}
       \caption[Sample images]
      {Sample $\rmag$-band images. We present our visual classification based on the Hubble classification in the top-left corner of each image: ellipticals (E); lenticulars (S0); early spirals (S$_{\rm E}$); and late spirals (S$_{\rm L}$). Each image covers three times $\overline{r_{90}}$ of each galaxy.}
       \label{hcat}
     \end{figure}

     \begin{figure}
       \centering
       \includegraphics[width=0.5\textwidth]{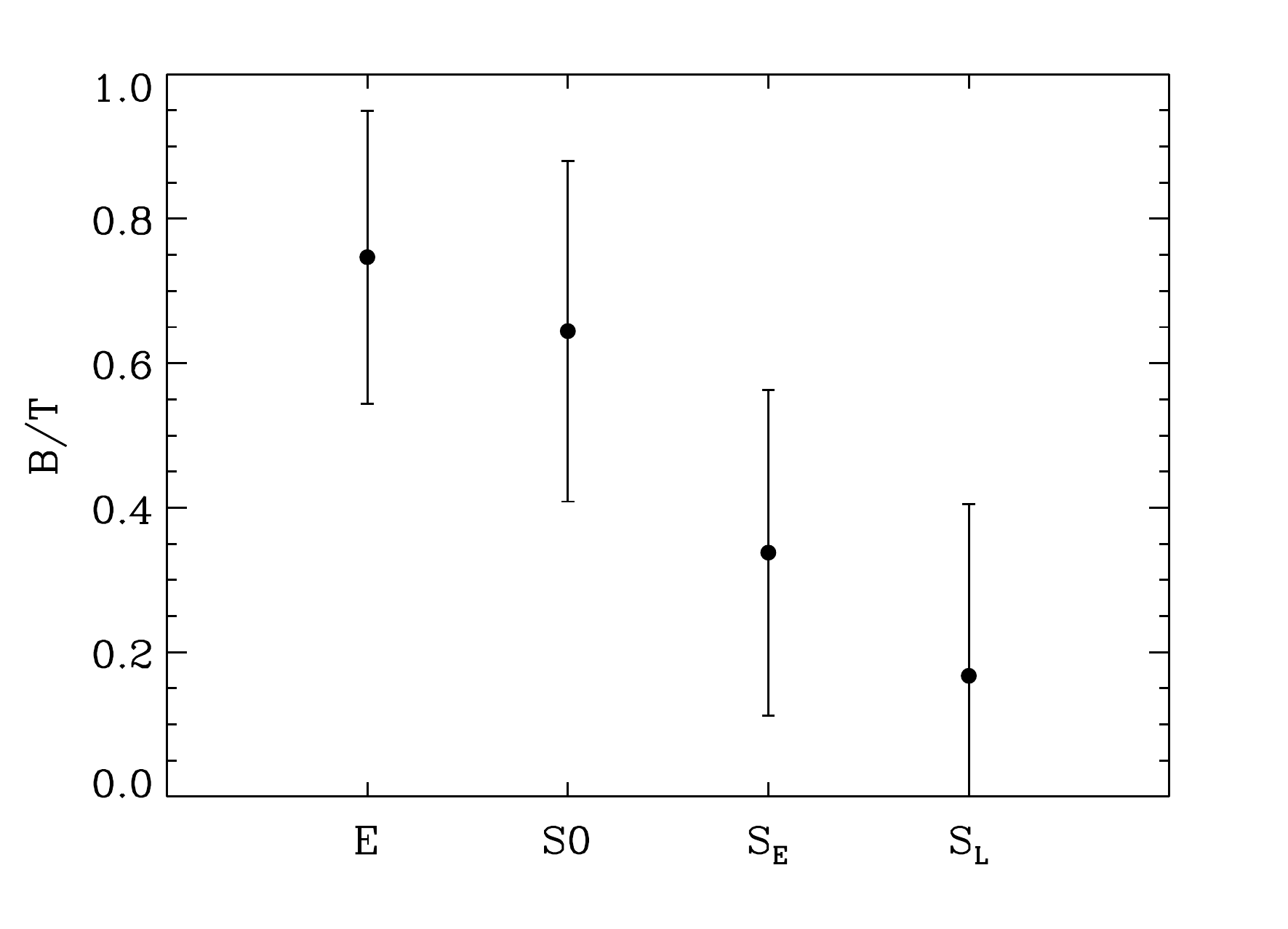}
       \caption[Hubble morphology and B/T from {\sc{galfit}}.]
      {Hubble morphology and B/T from {\sc{galfit}}. We present mean B/T value of each morphology. The error bars show $68\%$ probability distributions for each morphology. The morphology which is determined through visual inspection is reasonably distributed across B/T. }
       \label{morbt}
     \end{figure}

   \subsubsection{Merger-related disturbance identification}
Galaxies showing evidence of galaxy mergers were identified from galaxies without any features (N) by focusing on whether they show any disturbance in morphology including one or more following features: merger features, multiple cores, tidal features, shell structures, unusual dust lane, and asymmetric light distribution. Example cutout image set for each feature is shown in Figure~\ref{appmer} in Appendix~\ref{app:vis}.
    
\begin{figure}
       \centering
       \includegraphics[width=0.45\textwidth]{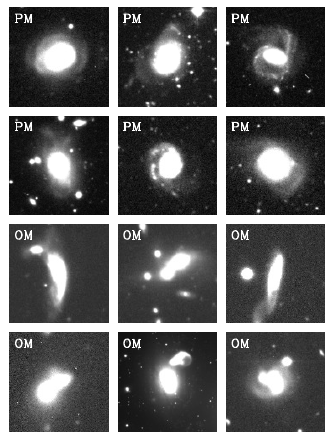}
       \caption[Sample images of galaxies with post-merger features]
      {Sample $\rmag$-band images of galaxies showing visual evidence of ongoing mergers (OM) and post mergers (PM). Each image covers three times $\overline{r_{90}}$ of each galaxy.}
       \label{mcat}
     \end{figure}

We divided galaxies showing disturbed features into two subgroups according to the existence of their companions: galaxies with features caused by the ongoing mergers with close pairs within four times $\overline{r_{90}}$ (OM); and galaxies with features seemingly related to mergers in the past without relevant pairs for the features (PM). Figure~\ref{samvis} (a) was useful for identifying OM by examining the existence of counterparts for the merger-related features. We present sample images of OM and PM in Figure~\ref{mcat}.
 
\begin{figure}
       \centering
       \includegraphics[width=0.5\textwidth]{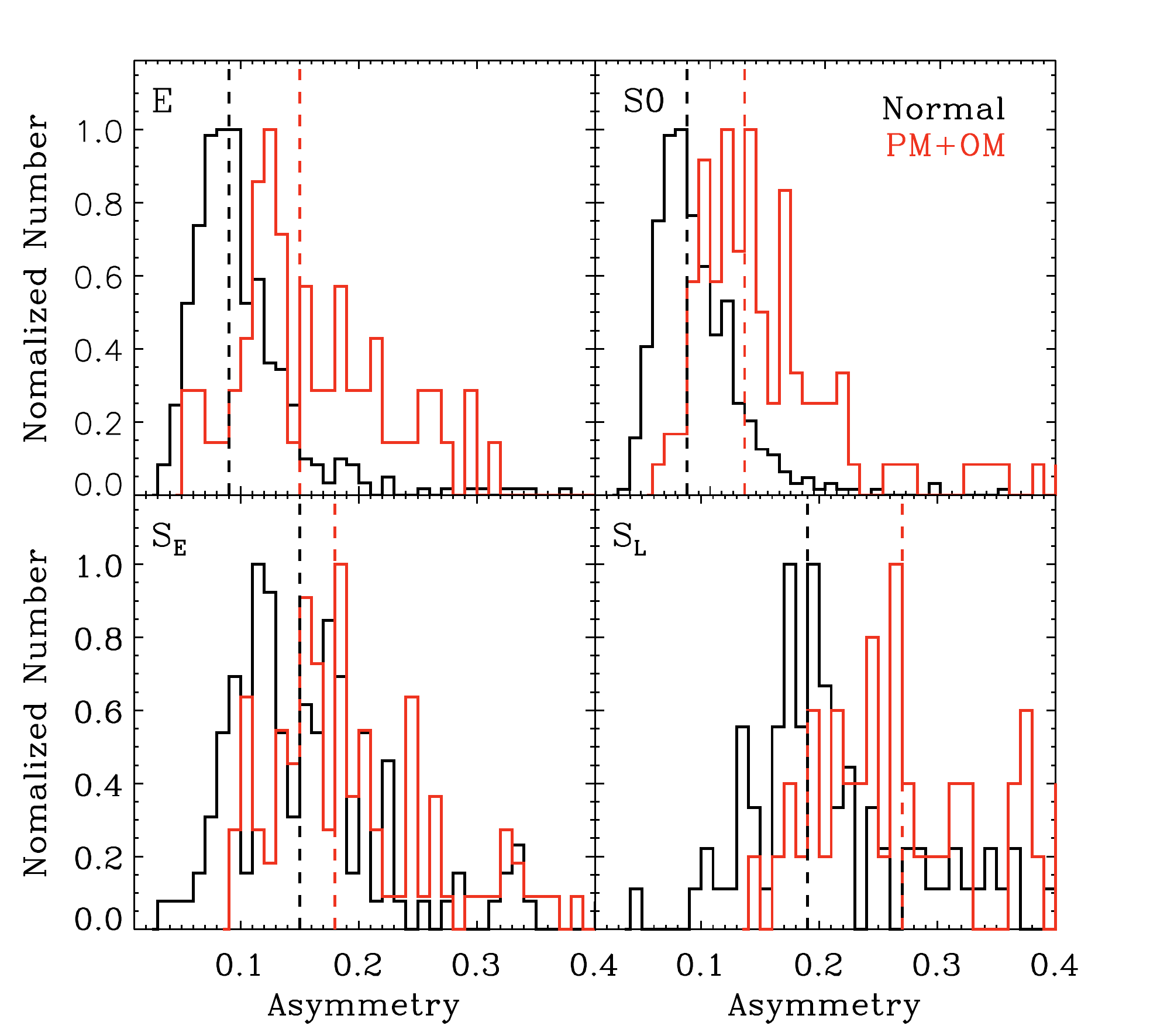}
       \caption[Asymmetry]
      {The CAS asymmetry for each morphological type. Dashed lines show the median value of asymmetry. Normal galaxies (black) have a smaller value of asymmetry than merger-related groups (red; PM and OM).}
       \label{asym}
     \end{figure}

The identification of merger-related disturbance features is admittedly subjective, having a chance to be contaminated by other environmental interactions. Ram-pressure stripping making features along the direction of the path is one of the possible contaminations. However, our post-merger galaxies tend to have asymmetric features which are distinct from the features generated by ram-pressure stripping. We measured the CAS asymmetry parameter on the $\rmag$-band deep images as a sanity check of our classification (Abraham et al. 1996; Conselice 2003). We limited the measurement window to 1.5 times the radii containing 90 percent of the total luminosity to reduce contaminations. Merger-related galaxies (PM and OM) have a distinct distribution of the asymmetry parameter compared with that of normal galaxies although the distributions are not separated sufficiently to identify merger-related galaxies based solely on the asymmetry parameter (Figure~\ref{asym}). Late-type galaxies (S$_{\rm E}$ and S$_{\rm L}$) show a larger value of asymmetry than early-type galaxies likely due to their spiral features which weaken the ability of asymmetry parameter to differentiate between normal and merger-related groups.

Our post-merger sample can be contaminated by flyby interactions. Flyby encounters are expected to generate similar features with minor mergers even without actual collisions (Sinha \& Holley-Bockelmann 2015). The frequency and outcome of flybys in the cluster environment are less informed and should be constrained by simulations. However, flybys can be considered subcategories of galaxy interactions. In this study, we assumed flybys bring a similar outcome with minor mergers.

\subsubsection{Spatial resolution}
Our sample clusters span a range of redshift (from 0.016 to 0.144), which generate a significant difference in spatial resolution between clusters. A coarse spatial resolution of clusters at higher redshift can blur disk structures and merger features, which might cause bias in our classification according to the redshift.

We randomly chose 170 galaxies from nine clusters at $z < 0.1$ (10 galaxies from Abell 3574 and 20 galaxies each from the others) and artificially degraded images by assuming the galaxies are at $z=0.144$ which is the largest redshift of our sample. Those degraded images were mixed to the original image set and have been classified by five researchers without knowing which ones are artificial images. Tables~\ref{tab:spa1} and \ref{tab:spa2} show the changes in the Hubble and merger classification between the original and degraded images, respectively. The classification which is based on spatially degraded images shows 6\% (11/170) and 5\% (9/170) difference with our final classification in Hubble and merger types, respectively. We could not find a clear tendency of the changes according to the morphological classes, types of mergers, and redshifts. Therefore, we estimate that our classification has an uncertainty around $5\pm3$\%, but this uncertainty is less biased to morphological classes, types of mergers, and redshifts.

 \begin{table}
\centering
\caption[Impact of spatial resolution]{Impact of spatial resolution on Hubble classification}
\begin{tabular}{l c c c c c c} 
\hline \hline
&& \multicolumn{4}{c}{Degraded Images}\\
\hline
	& & E & S0 & S$_{\rm E}$ & S$_{\rm L}$ & Total\\
	\hline
\multirow{ 4}{*}{Original Images}& E & 42 & 2 & 0 & 0 & 44 \\
&S0 & 2 & 65 & 3 & 0 & 70\\
&S$_{\rm E}$ & 0 & 1 & 35 & 1 & 37\\
&S$_{\rm L}$ & 0 & 0 & 2 & 17& 19\\
\hline
& Total & 44 & 68 & 40 & 18\\

 \hline
\hline
\end{tabular}
\label{tab:spa1}
\end{table} 
 
 \begin{table}
\centering
\caption[Impact of spatial resolution]{Impact of spatial resolution on merger classification}
\begin{tabular}{l c c c c c c} 
\hline \hline
&& \multicolumn{3}{c}{Degraded Images}\\
\hline
	& & N & OM & PM & Total\\
	\hline
\multirow{ 4}{*}{Original Images}& N & 125 & 1 & 5 & 131 \\
&OM & 0 & 11 & 1 & 12\\
&PM & 2 & 0 & 25 & 27\\
\hline
& Total & 127 & 12 & 31 & \\

 \hline
\hline
\end{tabular}
\label{tab:spa2}
\end{table}

\section{Results}
\subsection{Morphological contents in galaxy clusters}
\begin{figure}
       \centering
       \includegraphics[width=0.5\textwidth]{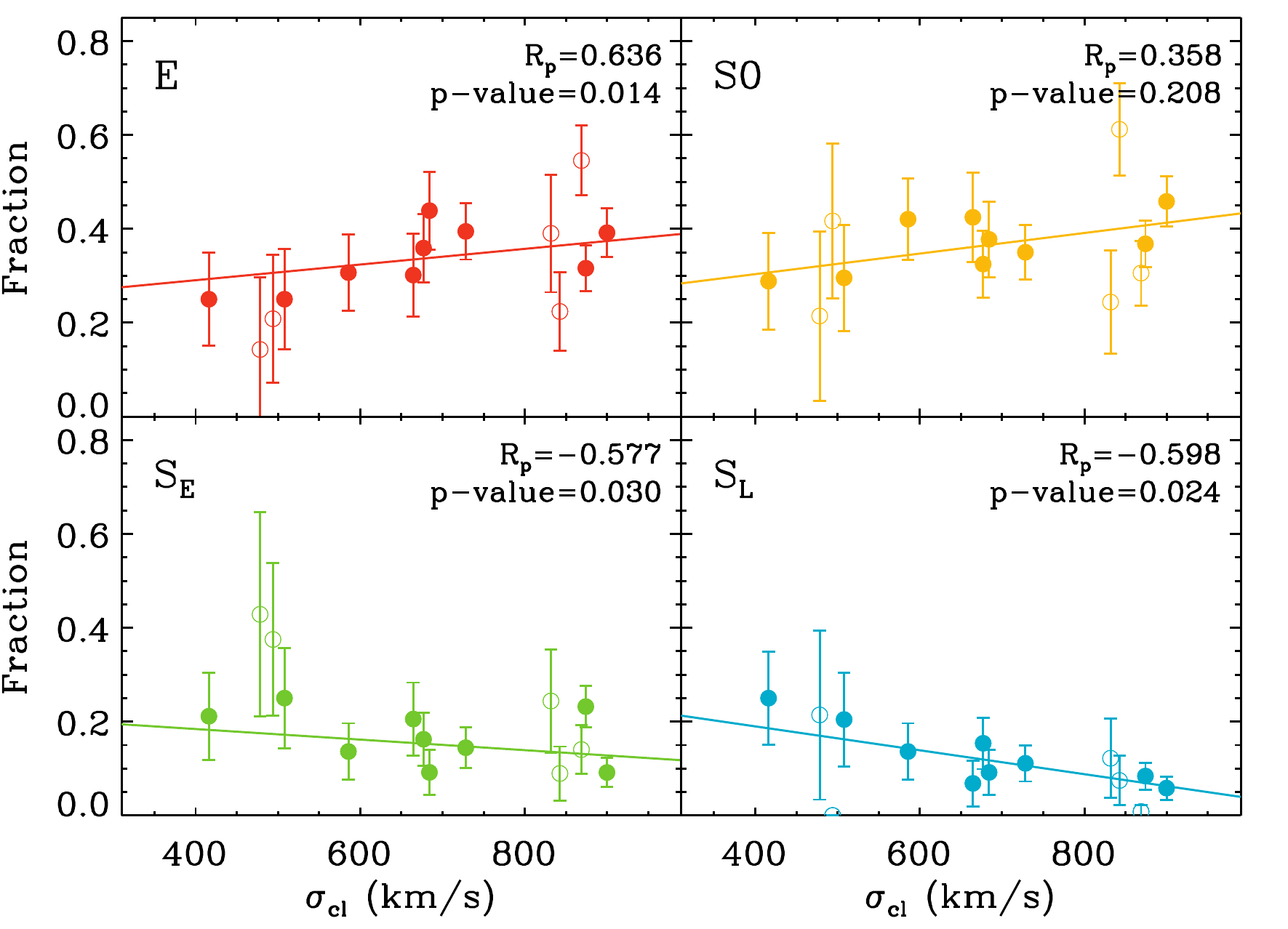}
       \caption[Hubble type and cluster velocity dispersion]
      {Morphological fractions based on visual inspection and cluster velocity dispersion. Open circles present clusters whose images do not cover $2\,R_{200}$ (Abell 1146, Abell 2589, Abell 3574, and Abell 3659) and Abell 1278 whose spectroscopic completeness is significantly low. The R coefficient ($R_{\rm p}$) and p-value from the Pearson correlation are shown in the top right corner of each panel. The error bar shows 90\% confidence interval for each fraction.}
       \label{morclt}
     \end{figure}

\begin{figure}
       \centering
       \includegraphics[width=0.5\textwidth]{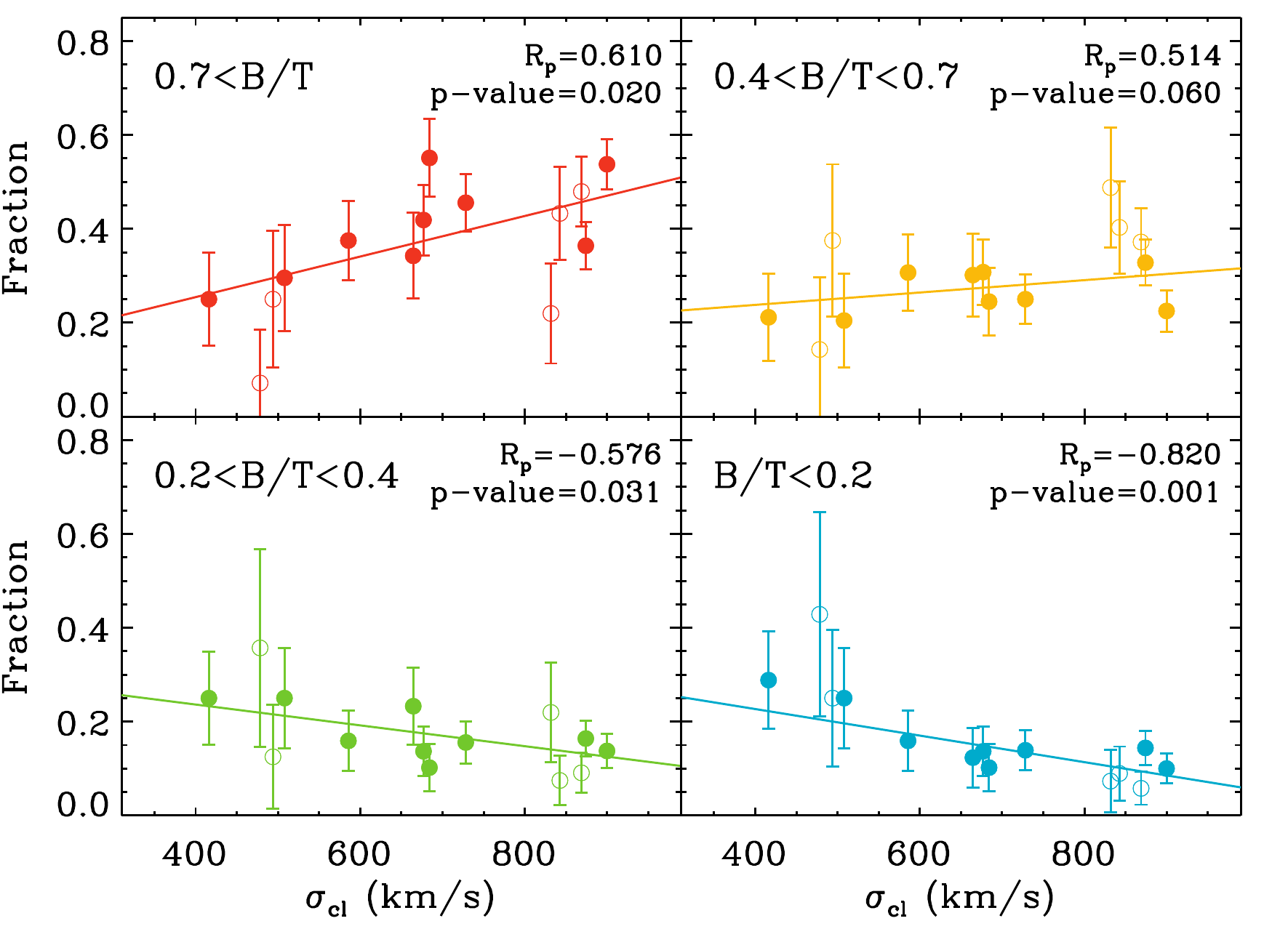} 
       \caption[Hubble type and cluster velocity dispersion]
      {Morphological fractions based on B/T ratio and cluster velocity dispersion. Open circles present clusters whose images do not cover $2\,R_{200}$ (Abell 1146, Abell 2589, Abell 3574, and Abell 3659) and Abell 1278 whose spectroscopic completeness is significantly low. The R coefficient ($R_{\rm p}$) and p-value from the Pearson correlation are shown in the top right corner of each panel. The error bar shows 90\% confidence interval for each fraction.}
       \label{btclt}
     \end{figure}

 We investigated galaxy morphology in our sample clusters as a function of the rest-frame cluster velocity dispersion ($\sigma_{\rm cl}$) which is a proxy for the cluster mass. The fraction of each morphology is clearly related to the cluster velocity dispersion (Figure~\ref{morclt}). Less massive clusters (e.g., $\sigma_{\rm cl}<600$ km s$^{-1}$) show a similar portion ($\sim$25\%) of all morphological types, whereas massive clusters (e.g., $\sigma_{\rm cl}>800$ km s$^{-1}$) are dominated by early-type galaxies. We also examined the correlation between cluster velocity dispersion and morphological contents based on B/T ratio from {\sc{galfit}}. In Figure~\ref{btclt}, we divided galaxies into four subgroups according to their B/T ratio: $0.7<B/T$; $0.4<B/T<0.7$; $0.2<B/T<0.4$; $B/T<0.2$. The fraction of B/T subgroups increases or decreases according to $\sigma_{\rm cl}$ in common with that of visual morphology. We employed the Pearson correlation to evaluate the statistical significance of the correlations. In Figures~\ref{morclt} and ~\ref{btclt}, the R coefficient ($R_{\rm p}$) and p-value from the Pearson correlation is presented in the top right corner of each panel. The p-value represents the probability of detecting a correlation by random shuffling, and therefore the lower the p-value, the more significant the correlation. The correlations between visual morphology and $\sigma_{\rm cl}$ have low p-values than 0.05 except for S0 galaxies. Also, we found low p-values in the correlations between the fraction of B/T subgroups and $\sigma_{\rm cl}$. Our results suggest that morphological contents of clusters based on both visual classifications and B/T ratios significantly correlate with cluster velocity dispersion of local clusters.

      \begin{figure}
       \centering
       \includegraphics[width=0.5\textwidth]{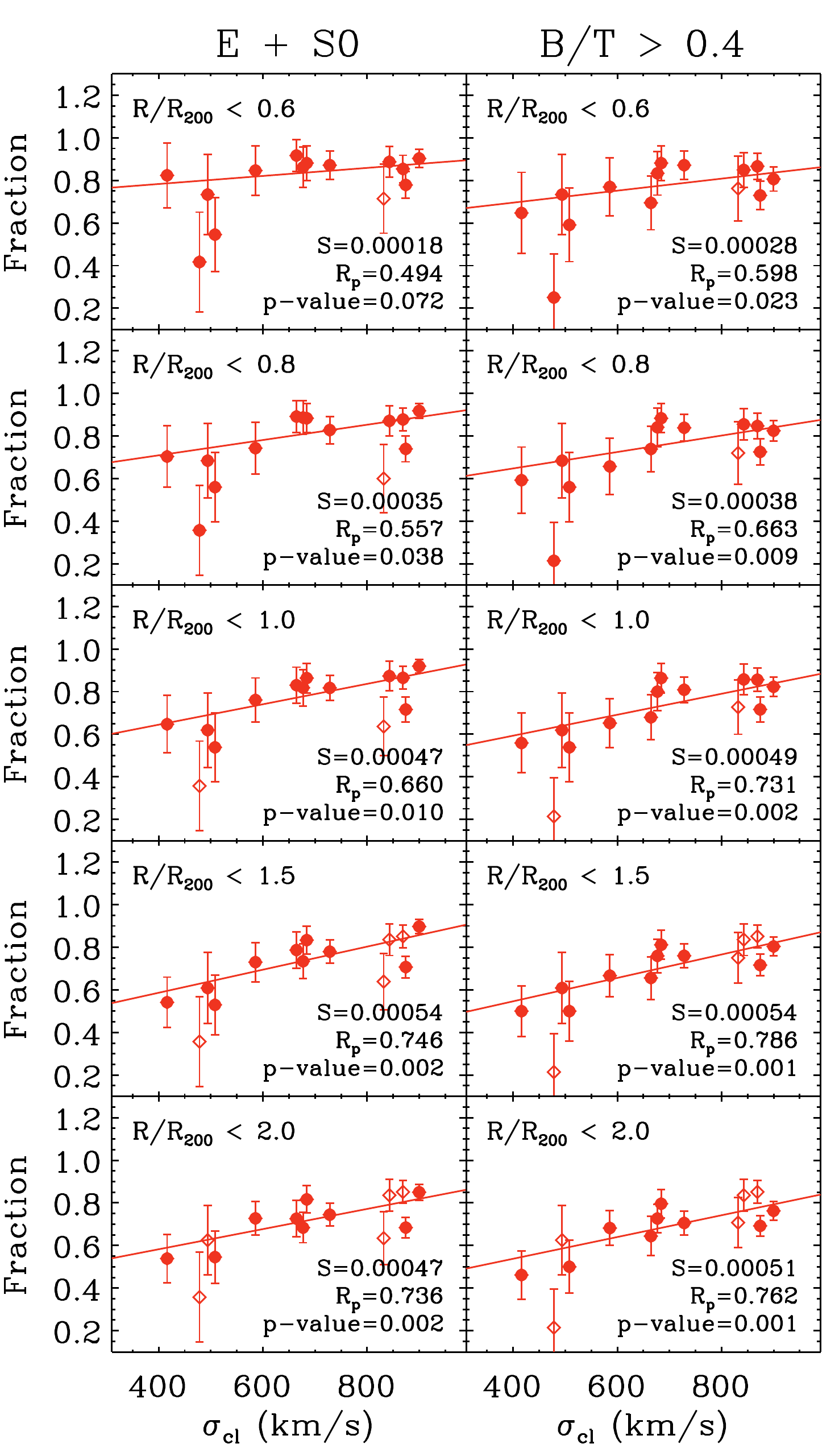}
       \caption[Hubble type and cluster velocity dispersion]
      {Fractions of early-type galaxies (left) and galaxies with B/T $> 0.4$ (right) to cluster velocity dispersion as a function of cluster-centric coverage. Open circles present clusters whose images do not cover the limiting cluster-centric distance or Abell 1278 whose spectroscopic completeness is significantly low. The slope (S), R coefficient ($R_{\rm p}$), and p-value from the Pearson correlation are shown in the bottom right corner of each panel. }
       \label{morcltdis}
     \end{figure}

 We investigated the fractions of early-type galaxies (E and S0) and galaxies having B/T$ > 0.4$ as a function of cluster-centric coverage in Figure~\ref{morcltdis}. Large values of the Pearson R coefficient (i.e., $R_{\rm p} > 0.7$) and small p-values (i.e., p-value $< 0.05$) indicate that both early-type and bulge-dominant galaxy fractions strongly correlate with cluster velocity dispersion especially for the sample covering large cluster-centric distance ($R/R_{200} > 1$). Moderate correlations are found within $R/R_{200} < 0.6$, and the correlation becomes stronger (large values of $R_{\rm p}$ and slope) and significant (small p-value) toward large cluster-centric coverage until $R/R_{200} = 1.5$. Almost identical trends were detected for the limiting coverage between $ 1.5 < R/R_{200} < 2$. These results imply that morphological transformations happen within the virialized region of clusters causing the dominance of early-type (or B/T $> 0.4$) galaxies in cluster core ($R/R_{200} < 0.6$), which dilutes the morphological dependence on cluster mass. 
 
A small sample size of the small geometrical coverage can also weaken the morphology-$\sigma_{\rm cl}$ correlation in Figure~\ref{morcltdis}, and the effect of sample size should be considered. For each geometrical coverage, we randomly selected the same number of galaxies at $R < 2R_{200}$ and measured $R_{\rm p}$ for the (E+S0)-$\sigma$ correlation (the left panel of Figure~\ref{morcltdis}). We repeated the random selection 1000 times and calculated the median values of $R_{\rm p}$ ($\overline{R_{\rm p}}$) and the standard deviation of the distribution ($\sigma_{\rm R_p}$). In Table~\ref{tab:rp}, ($R_{\rm p}-\overline{R_{\rm p}}$) shows the difference in $R_{\rm p}$ between geometrical and random selections. The $R/R_{200} < 1.5$ sample shows the similar $R_{\rm p}$ to that from the random selection and the difference is small (0.26 $\sigma_{\rm R_p}$). However, the small geometrical coverage samples ($R/R_{200} < 0.6$ or $R/R_{200} < 0.8$) show the 2-sigma smaller $R_{\rm p}$ (worse correlation) than that from the random selection ($\overline{R_{\rm p}}$). Therefore, we can conclude that small geometrical coverages indeed weaken the morphology-$\sigma_{\rm cl}$ correlation.
      
\begin{table}
\centering
\caption[The Pearson R coefficient distribution from random selection]{ The $R_{\rm p}$ distribution from the random selection (E+S0) }
\begin{tabular}{c c c c c c c} 
\hline \hline
 \multicolumn{3}{c}{Geometrical Selection}& \multicolumn{2}{c}{Random Selection}\tablenotemark{a}&\\
\hline
Coverage & {$N_g$}\tablenotemark{b}& $R_{\rm p}\tablenotemark{c}$ & $\overline{R_{\rm p}}\tablenotemark{d} $ & $\sigma_{\rm R_p}\tablenotemark{e}$ & ($R_{\rm p} - \overline{R_{\rm p}}$)\\
	\hline
 $R/R_{200}<0.6$& 688 & 0.494 & 0.696 & 0.106 & -1.91$\sigma_{\rm R_p}$ \\ 
 $R/R_{200}<0.8$& 861 & 0.557 & 0.716 & 0.070 & -2.27$\sigma_{\rm R_p}$ \\
 $R/R_{200}<1.0$& 978 & 0.660 & 0.728 & 0.056 & -1.21$\sigma_{\rm R_p}$ \\
 $R/R_{200}<1.5$& 1219 & 0.746 & 0.737 & 0.034 & 0.26$\sigma_{\rm R_p}$ \\
 \hline
\hline
\end{tabular}  
  \tablenotetext{1}{We randomly selected $N_g$ galaxies at $R<2R_{200}$ and repeated 1000 times.}  
  \tablenotetext{2}{Number of galaxies within the geometrical coverage.}  
  \tablenotetext{3}{The $R_{\rm p}$ from the $N_g$ geometrically-selected galaxies.}  
  \tablenotetext{4}{The median $R_{\rm p}$ from the random selection.} 
  \tablenotetext{5}{The standard deviation of $R_{\rm p}$ from the random selection.}  
\label{tab:rp}
\end{table}

Observational studies have reported the dependence of morphological contents on the cluster velocity dispersion at high redshifts (Postman 2005; Desai et al. 2007; Poggianti et al. 2009); however, it was not confirmed in the local universe. Poggianti et al. (2009) investigated the morphology of galaxies in clusters at $0.04<z<0.07$ based on the WIde-field Nearby Galaxy-cluster Survey (WINGS; Fasano et al. 2006) and could not find a clear correlation between the morphological fraction and the cluster velocity dispersion. Likewise, Simard et al. (2009) reported that there is no trend between the fraction of early-type galaxies and the cluster velocity dispersion based on local clusters from the SDSS data. 

These earlier findings on the morphological dependence on the cluster velocity dispersion conflict with our results. We suspect that different sampling strategy caused the discrepancy on this issue. First, both Poggianti et al. and Simard et al. samples are limited to the galaxies which locate inside $0.6\,R_{200}$; whereas, we extended our sampling criterion to $2\,R_{200}$. Besides, Poggianti et al. sample is limited to the clusters whose cluster velocity dispersion is greater than 500 km/s. In Figure~\ref{morcltdis}, it is difficult to find a correlation with only $\sigma_{\rm cl}> $500km/s clusters even for the $R < 2R_{200}$ sample.

Low levels of spectroscopic completeness of previous studies might also have weakened the morphology-$\sigma_{\rm cl}$ correlation. Following Cava et al. (2009), the spectroscopic completeness of WINGS clusters is around 0.5 which is considerably smaller than that of this study (Table~\ref{tab:comp}). We presume that the spectroscopic completeness of Simard et al. (2009) is also significantly lower than this study. For the sample galaxies within the SDSS footprint, the spectroscopic completeness based only on the SDSS data is 36\% (1276/3577) which is much smaller than ours (71\%; 2547/3577). Spectroscopically incomplete samples have more chance to be biased toward specific types of galaxies (usually toward bright galaxies), which can generate a bias to the morphological fraction of clusters. Also, spectroscopic incompleteness can bring a significant error on the estimation of cluster velocity dispersion.



\subsection{Morphology-density relation}
\label{sec:den}
Having spectroscopic information of all the galaxies in the field allows one to accurately measure local densities.
In this study, however, the spectroscopic completeness reaches 1 in only two clusters (Table~\ref{tab:comp}). The measurement based on projected distance highly overestimates local density due to foreground/background contaminations. For a more realistic measurement of local density, we followed the method described in Fasano et al. (2015). First, we estimated the circular area per Mpc ($a_i$) that includes the $i$th nearest neighbor by the following equation:
\begin{equation}
N_{i} = i/f_c - N_F \,a_i,
\end{equation}
where $f_c$ is the imaging coverage fraction for $a_i$, and $N_F$ is the number of contaminations per Mpc. In most cases, $a_i$ is fully covered by our image, and therefore $f_c$=1. However, local density of galaxies on the edge of the image can be overestimated because we cannot count galaxies beyond the FOV. In such cases, we calculated $f_c$ by dividing the covered area by $a_i$. Out of 1409 total samples, 51 galaxies have $f_c$ smaller than 1. We adopted $N_F$ from Berta et al. (2006) who counted the number of field galaxies per square degree and per V-band magnitude based on the ESO-Spitzer wide-area Imaging Survey (ESIS) data. The V-band magnitude in their study was transformed into the $\rmag$ magnitude by utilizing the transformation formula in Fukugita et al. (2007). We calculated $a_i$ by increasing $i$ until we get $N_{i}$ greater than 10. Then, we calculated the area including the 10th nearest neighbor ($A_{10}$) by interpolating $a_i$ and $a_{i-1}$. Finally, we computed local density ($\Sigma_{10}$) using the 10th nearest neighbor as follows:
\begin{equation}
{\rm log}\,\Sigma_{10} = 10 / A_{10} + 0.178\,(f_c-1).
\end{equation}
The term of 0.178\,($f_c$--1) comes from Fasano et al. (2015) by comparing local density with the result, including the SDSS galaxies which locate outside their FOV. 

     \begin{figure}
       \centering
       \includegraphics[width=0.5\textwidth]{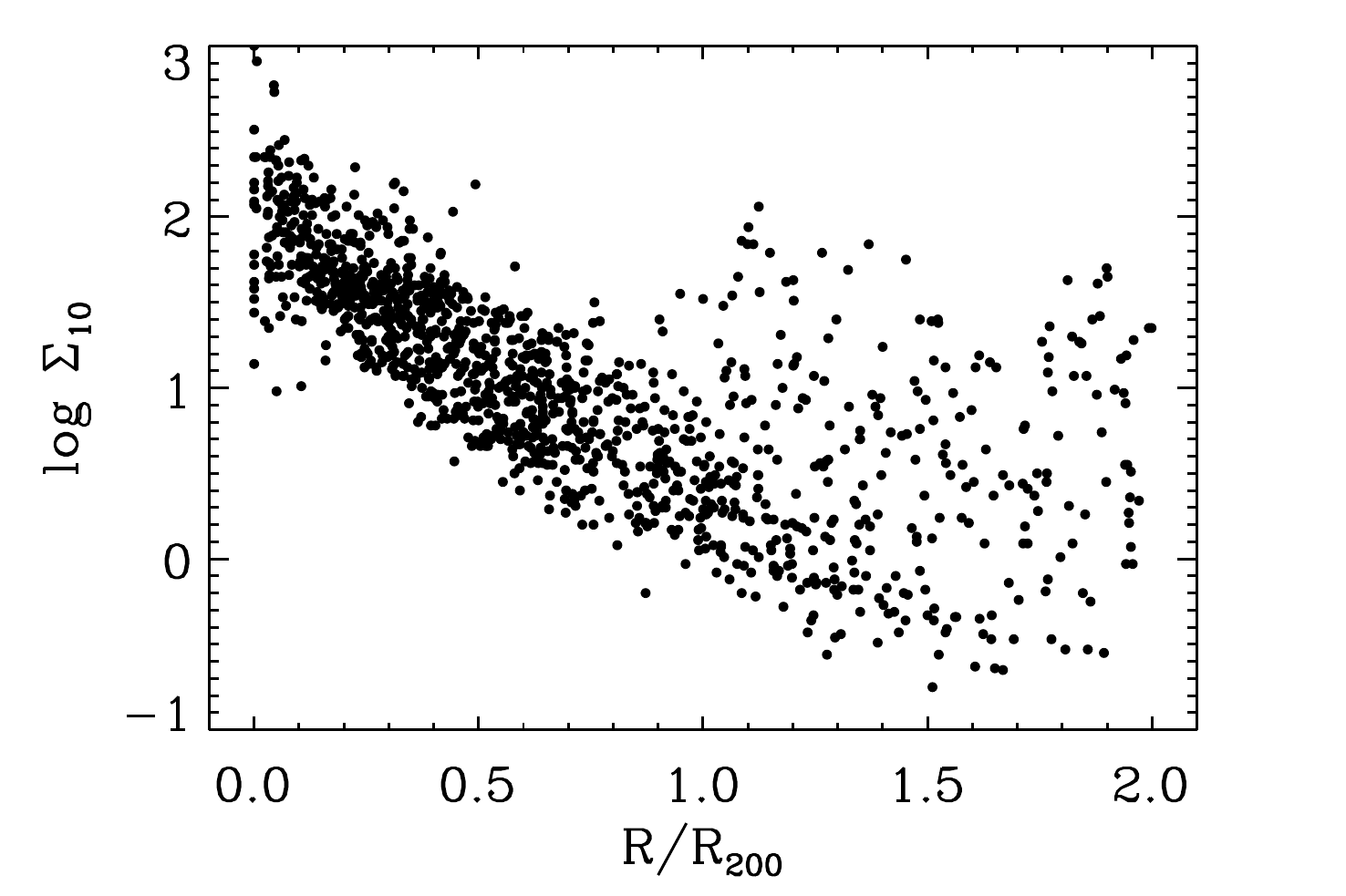}
       \caption[The density and the cluster-centric distance]
      {The density and the cluster-centric distance. Two parameters are well correlated to each other until the virial radius ($R/R_{200} = 1$). The outskirt regions of clusters ($R/R_{200} > 1$) show large scatter between two parameters.}
       \label{disden}
     \end{figure}

There is a correlation between local density and the cluster-centric distance (Figure~\ref{disden}). The cluster center shows the highest local density, and local density decreases as the distance from the cluster center increases. Within the virial radius ($R/R_{200}<1$), local density shows a tight correlation with $R/R_{200}$. Although the outskirt region of clusters ($R/R_{200}>1$) tends to show a smaller value of local density, the large scatter in $\Sigma_{10}$ indicates that significant numbers of galaxies in $R/R_{200}>1$ are located in locally dense regions.

     \begin{figure}
       \centering
       \includegraphics[width=0.5\textwidth]{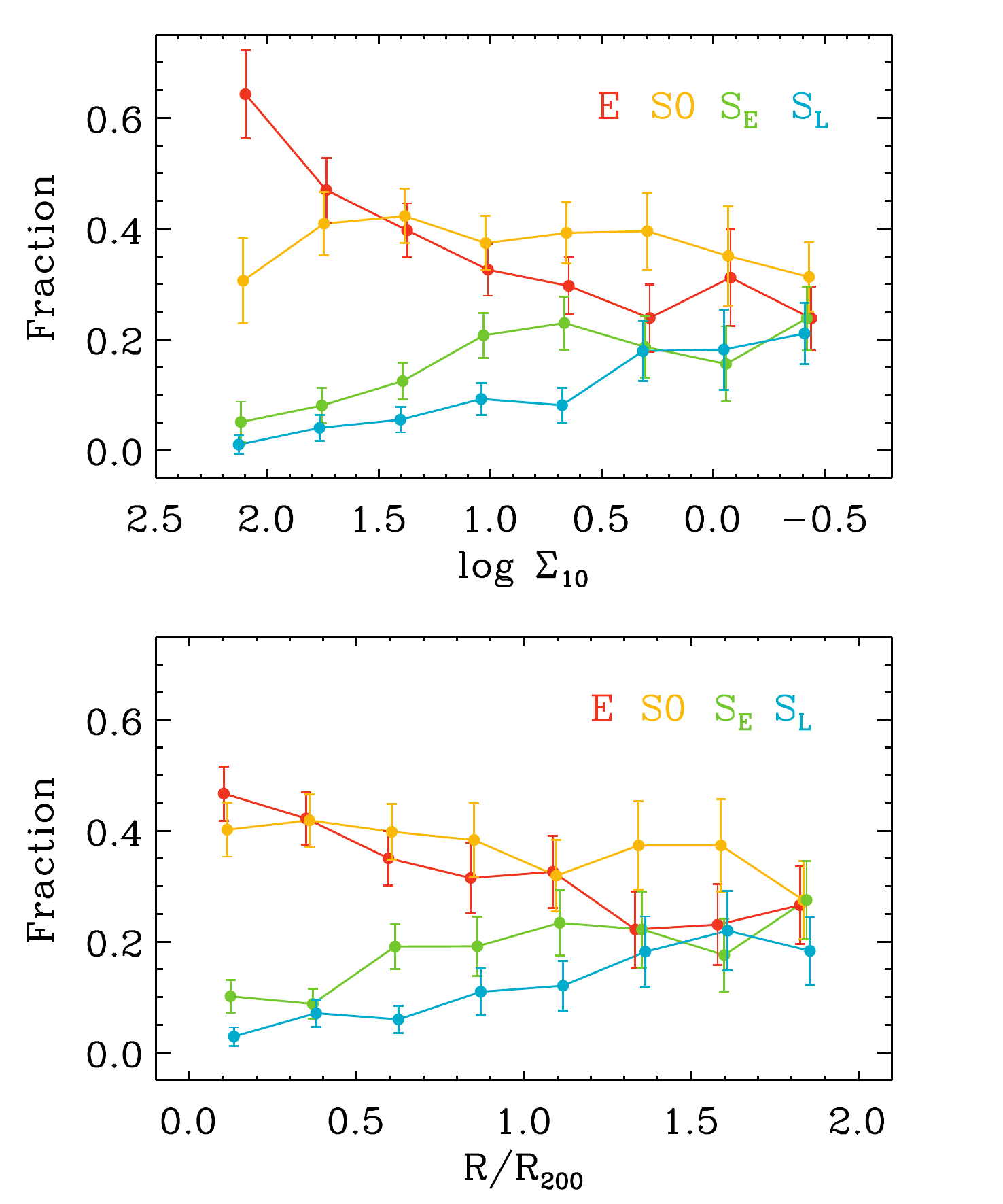}
       \caption[The morphology-density relation]
      {The morphology-density (top) and morphology-radius (bottom) relations in the cluster environment. The error bar shows 90\% confidence interval for each fraction. The fraction of elliptical galaxies increases with local density or cluster-centric distance; whereas, the fraction of late-type galaxies decreases with local density comparatively. The fraction of lenticular galaxies does not change significantly with local density.}
       \label{morden}
     \end{figure}

We present the morphology-density relation based on cluster galaxies in Figure~\ref{morden}. Highly dense regions preferentially have E and S0 galaxies. The lower local density, the lower the E fraction and the higher the S$_{\rm E}$ and S$_{\rm L}$ fractions. The correlation between local density and cluster-centric distance generates similar results on the morphology-density and the morphology-radius relations. Smaller values of the cluster-centric distance increase the fraction of early-type galaxies. As a result, we barely find S$_{\rm E}$ or S$_{\rm L}$ galaxies in the central region of clusters. The changes in the S0 fraction against local density (or cluster-centric distance) are not significant. The overall trend of our cluster morphology-density relation is consistent with previous studies. Dressler et al. (1997) including a revised version of the Dressler (1980) morphology-density relation shows that elliptical and spiral fractions are well correlated to local density, whereas S0 fractions have relatively little dependence on local density within the cluster environment. Fasano et al. (2015) also reported a relatively constant fraction of S0s to local density based on WINGS clusters.

We found around 40\% of S0 galaxies at cluster outskirt region ($R/R_{200} > 1$) before they gravitationally bound to the cluster potential. Considering the low fraction of S0s in field environments ($\sim 24\%$; e.g. Calvi et al. 2012), we presume that the major morphological transformations from spirals to S0s have happened in the intermediate environments, somewhere between clusters and fields.

The group environment can be a possible place where this transformation happens (pre-processing; e.g. Mihos 2004). Previous studies suggest that group environments also hold the similar environmental effects with clusters such as strangulation, tidal interactions, and ram-pressure stripping which is considered one of the main mechanisms to trigger morphological transformations by removing gas in the outer disk (Rasmussen 2006; Hester 2006). However, the effect of ram-pressure stripping in the group environment is much milder than that in massive clusters (Christlein \& Zabludoff 2004; Rasmussen et al. 2008), and therefore, it is considered inefficient for triggering morphological transformations of large spirals. Freeland, Sengupta, \& Croston (2010) also suggested that it is hard to explain the HI deficiency of group galaxies solely with ram-pressure stripping. Therefore, it is speculated that ram-pressure stripping is less likely the main mechanism for the morphological transformation occurred before the galaxies came to cluster environments.

If a galaxy-galaxy merger is the most effective way for the morphological transformation, it could have preferentially occurred in group environments rather than in cluster environments (Barnes \& Hernquist 1992). Just et al. (2010) reported that S0 fractions have rapidly increased in groups or poor clusters since $z \sim 0.5$ but not in rich clusters suggesting that galaxy-galaxy interactions are responsible for the present S0 galaxies. The merger-induced S0 formation scenario also explains a non-negligible fraction of S0 galaxies in field environments.

Nearly constant S0 frequency over the cluster region does not necessarily rule out the morphological transformation in the cluster environment. Moreover, we still detect increasing/decreasing fractions of elliptical/spiral galaxies with respect to increasing local density or decreasing cluster-centric distance. Although we cannot pinpoint the mechanism causing morphological transformations in the cluster environment, our results suggest that transformations from spirals to S0s and from S0s to ellipticals would happen at a similar rate, and the S0 fraction remains the same.

\subsection{Environmental quenching}
Ultraviolet fluxes are sensitive to young stars and often used to identify galaxies hosting recent episode of star formation (RSF; Yi et al. 2005; Kaviraj et al. 2007; Jeong 2009; Sheen et al. 2016). We tried to get a hint of environmental quenching on cluster galaxies based on the optical and near-ultraviolet (NUV; $\lambda_{\rm eff} = 2267\AA$) fluxes obtained from this study and the Galaxy Evolution Explorer (GALEX; Martin et al. 2005) GR 6 plus 7 Data Release.
  
      \begin{figure}
       \centering
       \includegraphics[width=0.5\textwidth]{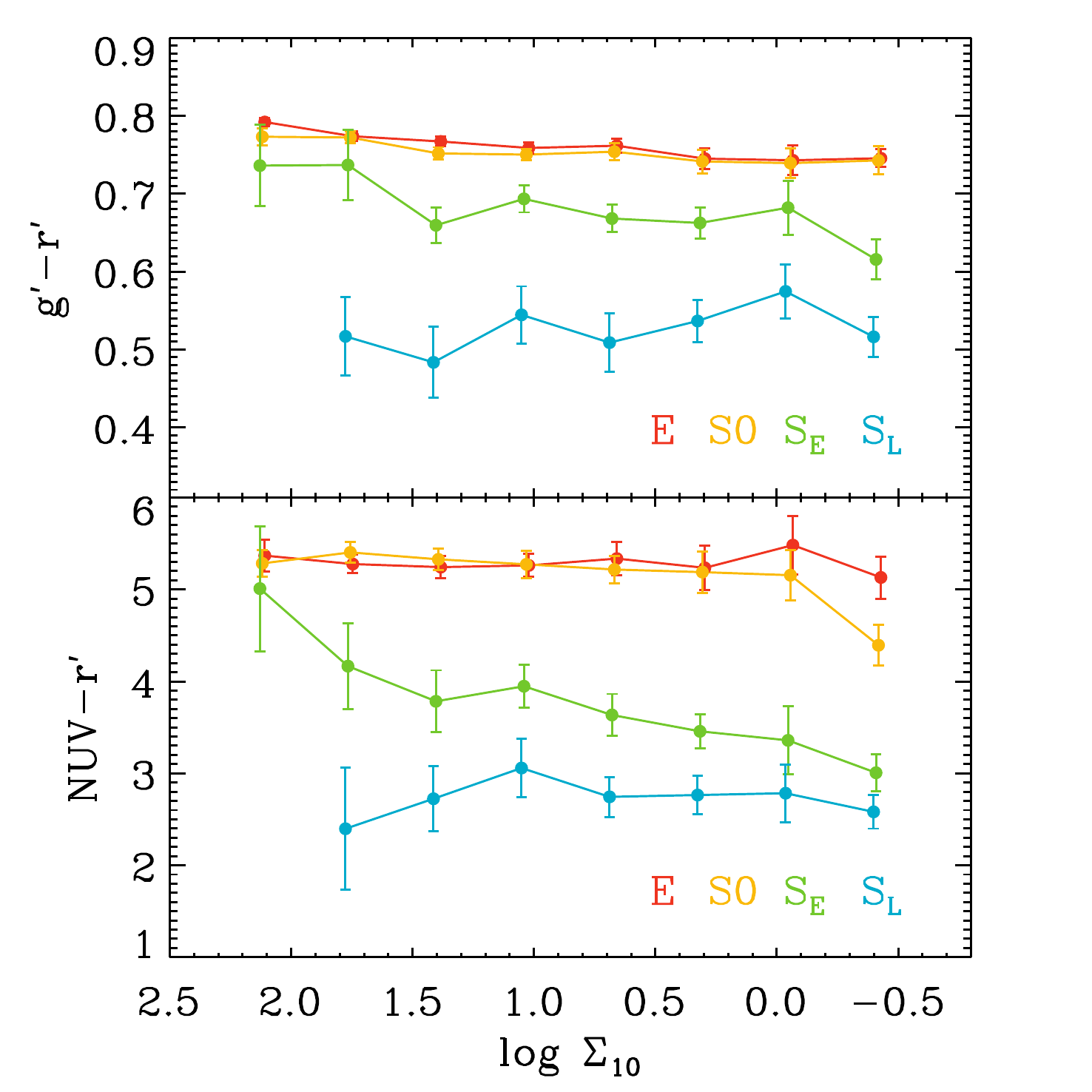}
       \caption[Galaxy colors and local density]
      {The $\gmag-\rmag$ (top) and NUV-$\rmag$ (bottom) colors as a function of local density. The error bar represents the standard error of the median. The optical and NUV colors show a similar trend with local density. The colors of S$_{\rm E}$ types significantly change according to local density; whereas the colors of the other types do not change with local density. }
       \label{dencol}
     \end{figure}
      
\begin{figure}
       \centering
       \includegraphics[width=0.5\textwidth]{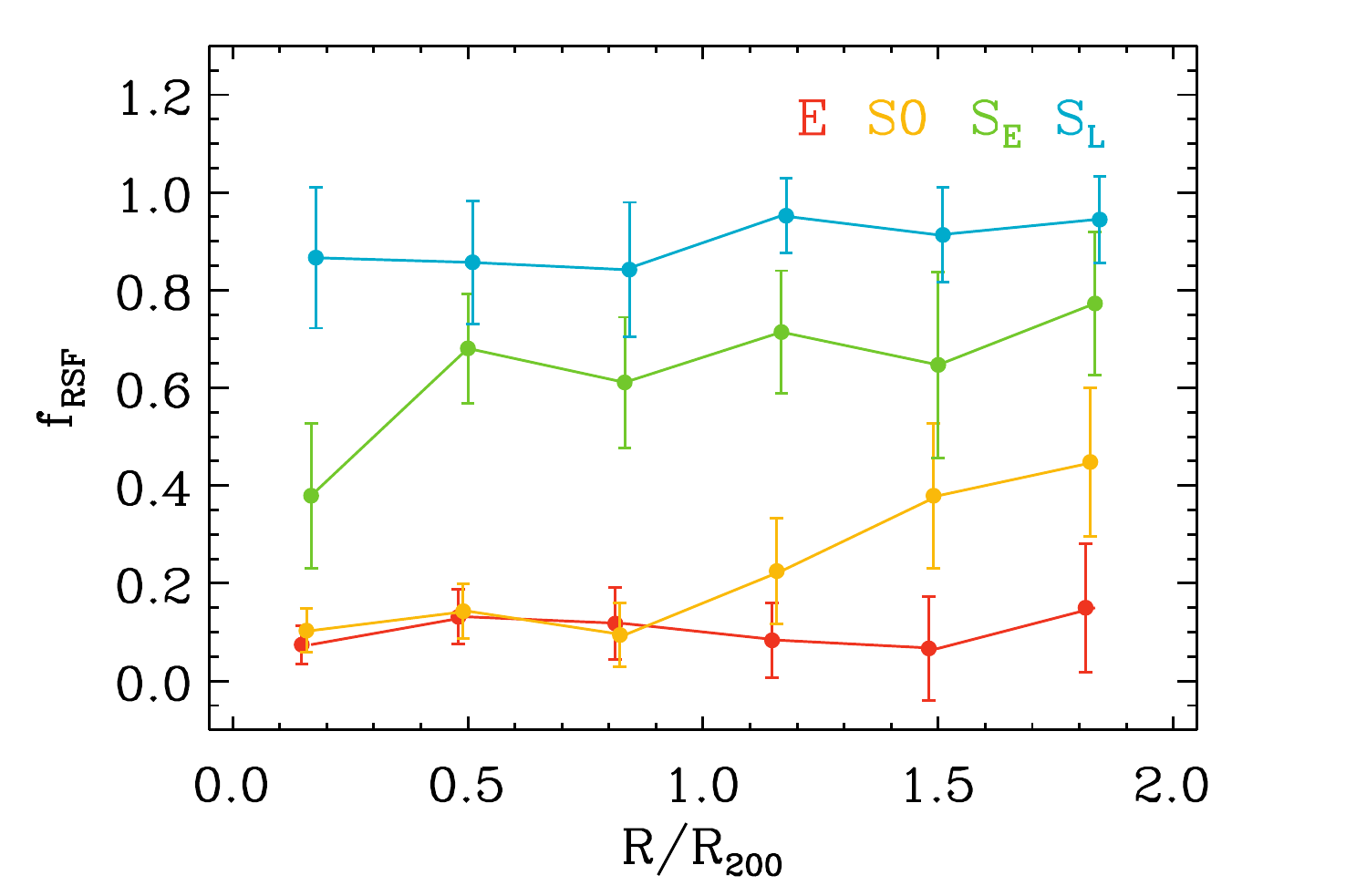}
       \caption[Galaxy colors and local density]
      {The fraction of galaxies showing recent star formation ($f_{\rm RSF}$) as a function of cluster-centric distance. The error bar represents 90\% confidence interval for each fraction.}
       \label{rsf}
     \end{figure}

In Figures~\ref{dencol} and~\ref{rsf}, we present the median $\gmag-\rmag$ and $NUV-\rmag$ colors and the RSF galaxy fraction ($f_{\rm RSF}$) of each morphological type as a function of local density or cluster-centric distance, respectively. The RSF galaxies are conservatively defined by selecting galaxies whose $NUV-\rmag$ lower than 4 corresponding to the value 2$\sigma$ below from the red sequence of E types in the $NUV-\rmag$ color-magnitude diagram.

 The median colors of late-type galaxies are much bluer than that of early-type galaxies even in the cluster environment. S$_{\rm L}$ type galaxies are dominated by RSF galaxies ($f_{\rm RSF}\sim0.9$) maintaining their blue color regardless of local density, although we barely find S$_{\rm L}$ type galaxies in the highly dense region (i.e., cluster center). On the other hand, S$_{\rm E}$ types show a significant change in colors according to local density especially at dense region (e.g. ${\rm log}\,\Sigma_{10} > 1.5$ or $R/R_{200} < 0.5$) where we found a sudden decrease in $f_{\rm RSF}$. This finding is in agreement with Solanes et al. (2001) who reported that HI-deficient fraction of early spirals significantly increases toward cluster center, but that of late spirals rather mildly increases. We could not find a statistically significant difference in the mean mass and B/T according to local density within the same morphological type. Therefore, the color change in S$_{\rm E}$ types is less likely to come from the difference in the mass or bulge fraction.
   
  Late-type galaxies actively forming stars seem to lose their gas in the cluster potential. The cosmological hydrodynamic simulation by Cen, Pop, \& Bahcall (2014) suggested that gas-rich galaxies lose most of their cold gas during the first infall into the cluster center. In that sense, a significant number of S$_{\rm E}$ types would have spent a long time in the cluster environment losing their gas. As a result, we found a significant change in colors and $f_{\rm RSF}$ of S$_{\rm E}$ at $R/R_{200} < 0.5$. On the other hand, it may not have been long enough for S$_{\rm L}$ types to lose their cold gas since their infall into the cluster environment.
   
Lenticulars hardly change their red colors and show low $f_{\rm RSF}$ within the virialized region (i.e., ${\rm log}\,\Sigma_{10} > 0.5$ or $R/R_{200} < 1$). However, we can find bluer $NUV-\rmag$ color and higher $f_{\rm RSF}$ from S0s at cluster outskirt region. Present cluster S0s would include both populations which already had S0 morphology even before they came into clusters (hereafter preformed S0s) and populations which were originally spirals when they arrived at cluster outskirts and transformed into S0s within cluster environments (hereafter transformed S0s). Most S0s at $R/R_{200} >1$ would be preformed populations, but S0s within the virialized region ($R/R_{200} < 1$) would be a mixture of both preformed and transformed populations, which makes it difficult to find the unique origin of (cluster) S0s.

For transformed S0s, we suspect that the star formation was quenched in a cluster before they changed their morphology from spirals into S0s, and therefore we do not detect bluer colors and higher $f_{\rm RSF}$ than Es within the virialized region. The changes in colors and $f_{\rm RSF}$ of S$_{\rm E}$ types also support this scenario, and therefore, we suspect that red S$_{\rm E}$ galaxies will eventually transform into S0s. The blue $NUV-\rmag$ color and higher $f_{\rm RSF}$ of preformed S0s at $R/R_{200} >1$ imply that their morphology transformation preceded star-formation quenching. That is, a significant fraction of preformed S0s may not be generated by a star-formation quenching derived by gas stripping. This explanation accords closely with our discussion on the constant S0 fraction at cluster outskirts in the previous section. Moreover, we found a link between galaxies with merger features and RSF galaxies, which will be discussed in the following series of this study based on the phase-space analysis by Rhee et al. (2017). 

Elliptical galaxies show red colors and low $f_{\rm RSF}$ ($\sim$ 0.1) no matter where they are located within a cluster. The star-formation quenching of transformed Es would precede morphological transformation because S0s at $R/R_{200} < 1$, possible progenitors of transformed Es, have already been quenched. Although preformed Es seem to have been quenched before they arrived in the cluster environment, it is difficult to speculate the quenching mechanism with our data. Observational and theoretical studies for the intermediate environment (i.e. groups) would give a clue to this question.

\subsection{Galaxy mergers in the cluster environment}
Galaxy clusters are considered a hostile environment for galaxy mergers, and we indeed found low frequency ($\sim$4\%) of the ongoing mergers. Note that the ongoing merger fraction is exactly matched with that in Sheen et al. (2012). However, around 20\% of cluster galaxies within $R/R_{200} < 2$ show signatures of post-mergers. In earlier work, Sheen et al. (2012) reported that 24\% of 167 red-sequence galaxies show signatures of post-mergers in 4 rich Abell clusters, whereas Adams et al. (2012) detected 3\% of post-merger galaxies among their 3551 early-type galaxies in the cluster environment. Even considering the subjectivity of the visual inspection, these two studies show contradicting results on the frequency of recent mergers in the cluster environment. The limiting surface brightness of Sheen et al. (2012) reaches $\mu_{\rmag}\sim$ 28 $\sur$, and Adams et al. (2012) and this study have used images with similar depth ($\mu_{\rmag}\sim$ 26.5 $\sur$ in 3$\sigma$). We can understand more detection of faint features in Sheen et al. (2012) because faint features can survive extremely long in low surface brightness (Ji, Peirani, \& Yi 2014). However, there is more than 15\% difference in the post-merger frequency between Adams et al. (2012) and this study. 

Sheen et al. (2012) and this study conducted a visual inspection using both original and model subtracted images; Adams et al. (2012), however, inspected only residual images after subtracting a model from $ellipse$ procedure. We experienced that $ellipse$ task sometimes fits faint features and diminishes the light related to disturbed features, making it hard to be detected. Moreover, Adams et al. (2012) examined only outer regions of galaxies to deal with extended light by masking out regions brighter than 1\% of the peak luminosity of a galaxy. In this study, we detect merger-related features (e.g., multiple cores, asymmetric light distribution) even near the galactic center. Besides, both Sheen et al. (2012) and this study utilized spectroscopically-confirmed cluster members; whereas Adams et al. (2012) selected cluster galaxies based on the red-sequence, which can contain foreground/background contaminations. These are possible explanations for the discrepancy on the detection of merger features, but we are still puzzled by such a big difference in the frequency of recent mergers in the cluster environment.

     \begin{figure}
       \centering
       \includegraphics[width=0.5\textwidth]{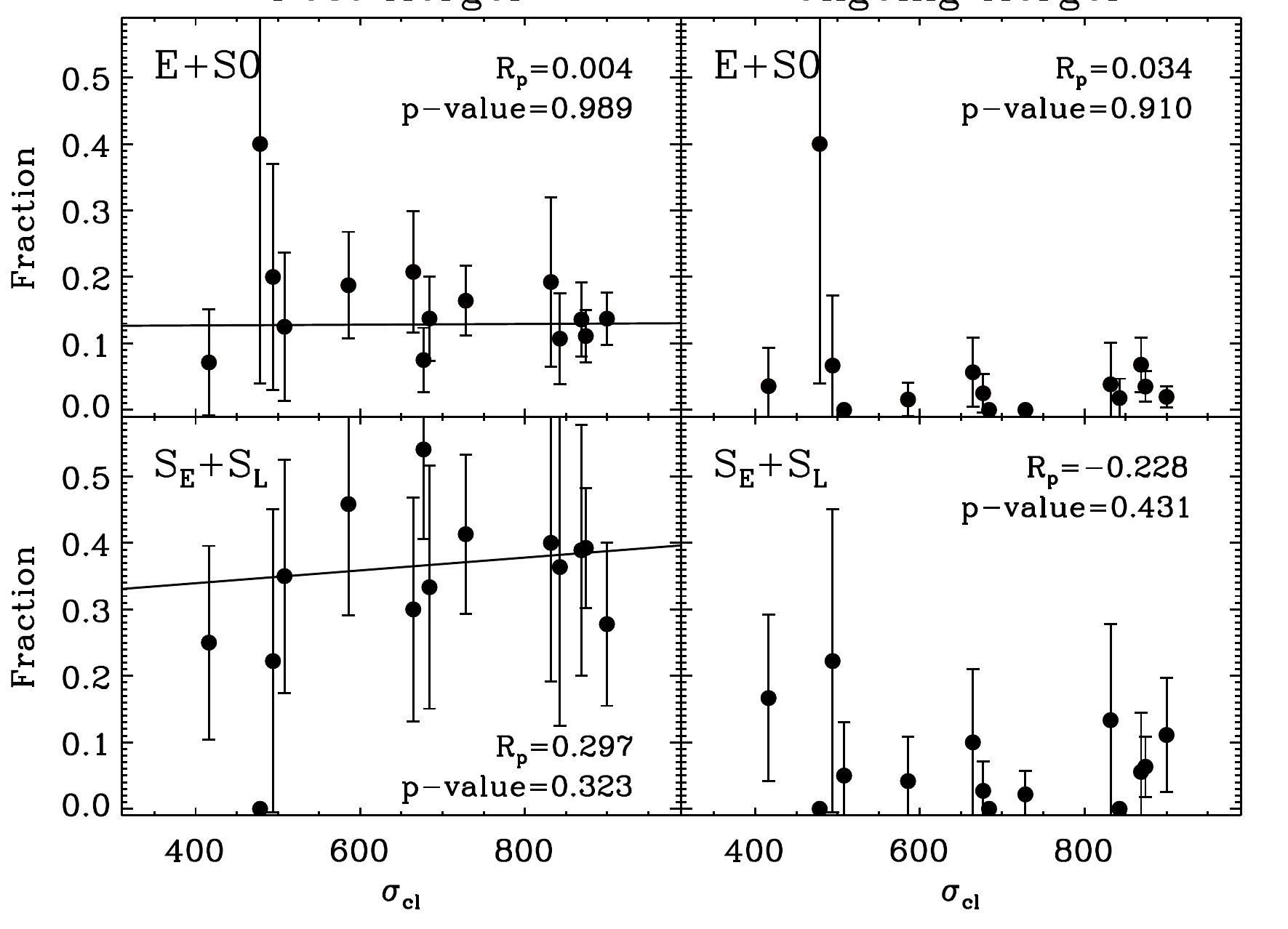}
       \caption[Galaxy mergers and cluster velocity dispersion]
      {Galaxy mergers and cluster velocity dispersion. The R coefficient ($R_{\rm p}$) and p-value from the Pearson correlation are shown in each panel. The error bar indicates 90\% confidence interval for each fraction. The cluster velocity dispersion and the frequency of galaxy mergers are not correlated each other. }
       \label{merclt}
     \end{figure}

We inspected whether galaxy mergers are preferentially found in more or less massive clusters (Figure~\ref{merclt}). We could not find a clear correlation between the frequency of galaxy mergers and the cluster velocity dispersion. We also detect a small fraction of ongoing mergers regardless of the cluster velocity dispersion. The large p-values from the Pearson correlation also support there is no relation between the frequency of mergers and cluster velocity dispersion. This result implies that the frequency of galaxy mergers is less correlated with the cluster potential. 

\begin{figure}
       \centering
       \includegraphics[width=0.5\textwidth]{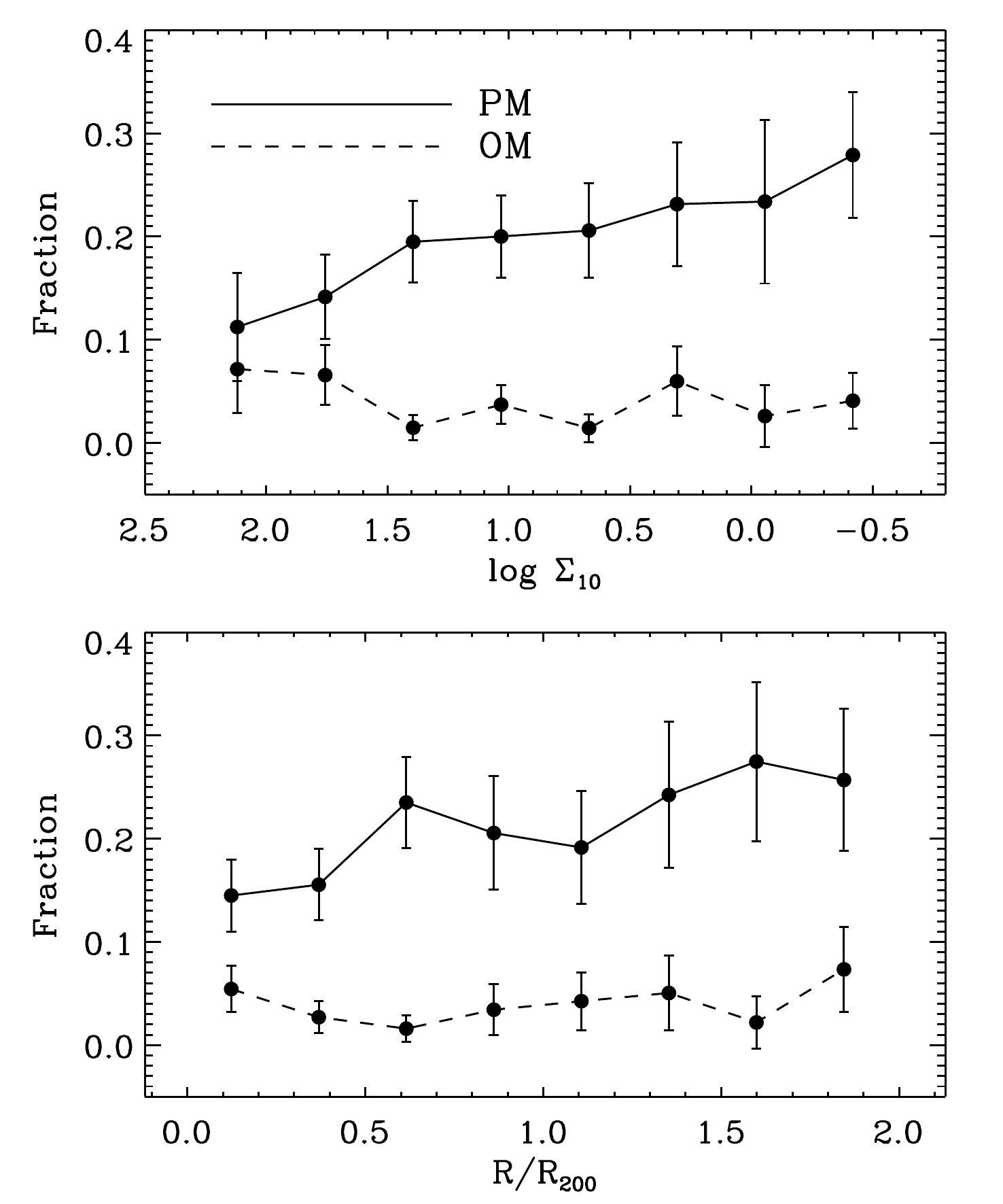}
       \caption[Galaxy mergers and local density]
      {The frequency of ongoing merger (OM) and post-merger (PM) groups along local density (top) and cluster-centric distance (bottom). The error bar shows 90\% confidence interval for each fraction. The fraction of post-merger decreases/increases with local density/cluster-centric distance; whereas, the fraction of ongoing merger remains constant with local density and cluster-centric distance.}
       \label{merden}
     \end{figure}

We examined the frequency of galaxy mergers according to local density and the cluster-centric distance (Figure~\ref{merden}). The PM fraction gradually decreases toward the dense region or the cluster center; whereas we cannot find clear tendencies on the detection rate of OM with local density. Our findings generate a speculation that the merger events which made the features we detect now have preceded the galaxy accretion into the cluster environment. The more frequent detection of PM in the cluster outskirt also support our scenario. 

The negative correlation between cluster-centric distance and the time since galaxies infall into the cluster environment explains the decreasing PM fraction toward the cluster center (Bah{\'e} et al. 2012; Rhee et al. 2017). Following Figure 3 in Bah{\'e} et al. (2012), most of the galaxies in cluster core (i.e. $R/R_{200} <$ 0.3) have spent 4 Gyrs or more time in cluster environments. Yi et al. (2013) numerically estimated the lifetime of merger features and suggested that merger features can survive 3.9 Gyrs in $\mu_{\rmag} \sim 28\,\sur$ depth. Although it is difficult to directly adopt the lifetime because it can vary depending on the type of mergers (e.g., mass ratio, merging orbit, and gas contents), we can expect smaller PM fraction in the central region than in outskirt region of clusters.

In our result, we found around five times higher fraction of PM galaxies ($\sim20\%$) than that of OM galaxies ($\sim4\%$). Following the hydrodynamic simulation by Ji, Peirani, \& Yi (2014), merger features can live 2.56 times longer than the merging time scale in a cluster potential assuming $\mu_{\rmag} \sim 28\,\sur$ (1.79 times for $\mu_{\rmag} \sim 25\,\sur$). The evolution of merger rate is another factor to be considered. Observational and theoretical studies presented that the merger rate in the past was higher than it is today (e.g., Bundy et al. 2009; Bridge, Carlberg, \& Sullivan 2010; Lotz et al. 2011; Lee \& Yi 2013; Rodriguez-Gomez 2015). Following the prediction by Rodriguez-Gomez (2015), the evolution of merger rate is proportional to $(1 + z)^{2.4-2.8}$ which suggest that merger rate declined roughly by a factor of 1.5 within the last 3 Gyrs. 

The low OM fraction (4.7\%) is difficult to explain the observed PM fraction (25.3\%) at cluster outskirts ($1.5 < R/R_{200} < 2$) even though we consider the evolution of merger rate and the timescale of merger features. The mergers made PM galaxies probably happened beyond $2R_{\rm 200}$ because of the time delay between OM and PM galaxies. We investigated PM and OM fractions beyond $2R_{\rm 200}$ using 207 galaxies which are brighter than -19.8 in the $\rmag$-band and have the similar redshift range to target clusters but locate beyond $2R_{\rm 200}$ and found 16 OM (7.7\%) and 56 PM (27.1\%) galaxies.

The significant difference in the detection of PM and OM in the cluster environment can be marginally explained by adopting all explanations regarding the evolution of merger rate, the timescale of merger features, and higher OM fraction beyond $2R_{\rm 200}$. However, our spectroscopic observations do not fully cover those fields beyond $2R_{\rm 200}$, and the spectroscopic completeness of $R>2R_{\rm 200}$ region is around 0.56 which is somewhat lower than that of $R<2R_{\rm 200}$ region (0.72). Besides, there are a lot of uncertainties in estimating timescales of mergers and features according to the details of mergers such as mass ratio, progenitor types, merging orbit, and environments. The PM and OM galaxies in the cluster environment can be fully explained through idealized observations targeting galaxies neighboring cluster regions (i.e., $R >2R_{\rm 200}$) and predictions of merging timescale from simulations for various types of mergers.

\section{Summary and Conclusions}
We carried out cluster deep imaging survey targeting 14 clusters at $0.015 \lesssim z \lesssim 0.144$ having a wide range of cluster mass. Finally, 1409 galaxies were selected as the member of clusters following the criteria: $|v_{\rm rv} - v_{\rm cen}| < 3\,\sigma_{\rm obs}$;  $R <2R_{200}$; $M_{\rmag} < -19.8$. We publish a value-added catalog (Table~\ref{tab:cat}) including photometry (Section~\ref{sec:phot}), redshift (Section~\ref{sec:red}), visual morphology (Section~\ref{sec:vis}), structural parameters from $ellipse$ task (Section~\ref{sec:ell}) and {\sc {galfit}} decomposition (Section~\ref{sec:gal}), and local density (Section~\ref{sec:den}). The first scientific series papers based on the data from the KYDISC analyzing pixel-based color-magnitude diagrams of BCGs and bright galaxies have been published recently (Lee et al. 2017, 2018). Studies of the impact of mergers on the evolution of cluster galaxies is reserved for later papers in this series.
  
We investigated morphological contents of sample clusters and found the dependence of galaxy population on cluster velocity dispersion. More massive clusters show more early-type galaxies, which suggests that galaxy clustering has affected the evolution of galaxies and changed their population. Even within a cluster, we can observe the relation between morphology and local density. The most distinct features in the cluster morphology-density relation is a nearly constant fraction of S0 galaxies. This implies that the major morphological transformation from spirals to lenticulars would be pre-processed before the galaxy accretion into the cluster environment (Mihos 2004).

We presume that present cluster galaxies within a virialized region ($R <R_{200}$) are the mixture of preformed populations which already had the present morphology even before they came into clusters and transformed populations which were transformed into the present morphological type within clusters, which makes it difficult to specify the evolution mechanism of cluster galaxies, especially for S0 galaxies. Cluster early-type galaxies (Es and S0s) are passive within a virialized region ($R <R_{200}$), which implies both that preformed early-type galaxies have already been quenched before they bound to the potential and that the quenching process within clusters possibly through gas stripping preceded morphological transformation to early-type galaxies. Some preformed S0s at $R >R_{200}$ show residual star formation, which makes us suspect that their morphology has been transformed before quenching star formation. We observed the evidence of environmental quenching in early spirals (S$_{\rm E}$). Some S$_{\rm E}$ types would have lived in a cluster for a sufficiently long time losing their gas. Late spirals (S$_{\rm L}$), however, have not spent enough time in a cluster to lose their gas and quench star formation.

From a small fraction of ongoing merger events ($\sim$ 4\%), we confirmed that galaxy clusters are hostile environments for galaxy mergers. However, we detect a considerable fraction of post-merger galaxies with visual signatures of recent mergers ($\sim$ 20\%). The post-merger fraction does not correlate with cluster velocity dispersion but depends on local density (cluster-centric distance). We conclude that galaxies with post-merger features have suffered the merger outside the cluster then fell into the cluster environment. Our results are consistent with Sheen et al. (2012) and the theoretical expectations from Mihos (2004), Fujita (2004), and Yi et al. (2013). 

Then, are galaxy mergers important to the evolution of cluster galaxies? Our findings suggest that galaxy mergers have less impact on galaxies which have been virialized to the cluster potential. Only recently accreted galaxies would have a chance to show merger-induced changes. In this regard, galaxy mergers have a limited influence on the present cluster population. However, the merger-driven evolution cannot be completely excluded when discussing the evolution of cluster galaxies. Galaxy clusters have accreted and will continue to accrete neighboring galaxies. Therefore, galaxy mergers affect cluster population in the way of pre-processing.

\acknowledgments
This study was performed under the umbrella of the collaboration between Yonsei University Observatory and the Korean Astronomy and Space Science Institute. S.K.Y. is the head of the project and acted as the corresponding author. Parts of this research were conducted by the Australian Research Council Centre of Excellence for All Sky Astrophysics in 3 Dimensions (ASTRO 3D), through project number CE170100013.
This paper includes imaging and spectroscopic data gathered with the 6.5-meter Magellan and the 2.5-meter du Pont Telescopes located at Las Campanas Observatory, Chile. This work made extensive use of imaging data based on observations obtained with MegaCam, a joint project of CFHT and CEA/IRFU, at the Canada-France-Hawaii Telescope (CFHT) which is operated by the National Research Council (NRC) of Canada, the Institut National des Science de l'Univers of the Centre National de la Recherche Scientifique (CNRS) of France, and the University of Hawaii. This work is based in part on data products produced at Terapix available at the Canadian Astronomy Data Centre as part of the Canada-France-Hawaii Telescope Legacy Survey, a collaborative project of NRC and CNRS.
This paper includes spectroscopic data based on observations at Kitt Peak National Observatory, National Optical Astronomy Observatory (NOAO Prop. ID: 2015A-0304 and 2015B-0315; PI: Sukyoung K. Yi), which is operated by the Association of Universities for Research in Astronomy (AURA) under a cooperative agreement with the National Science Foundation.

We are grateful to the anonymous referee for constructive comments and suggestions especially on the accuracy of photometry (Section 2.3), the effect of spatial resolution on the visual classification (Section 4.2.2), the impact of geometrical coverage on the Morphology-$\sigma_{\rm cl}$ relation (Section 5.1), and the constant S0 fraction (Section 5.2) that greatly improved the paper. We thank Matthew Colless for taking the time to give valuable advice on the merger frequency in the cluster environment.
S.K.Y. acknowledges support from the Korean National Research Foundation (NRF-2017R1A2A1A05001116). MK was supported by the Basic Science Research Program through the National Research Foundation of Korea (NRF) funded by the Ministry of Science, ICT \& Future Planning (No. NRF-2017R1C1B2002879). L.C.H. was supported by the National Key R\&D Program of China (2016YFA0400702) and the National Science Foundation of China (11473002, 11721303).  S.O. thank D.L. for the consistent support.

\appendix
\section{Sample Images}
 \label{app:vis}
  
     \begin{figure}
       \centering
       \includegraphics[width=1\textwidth]{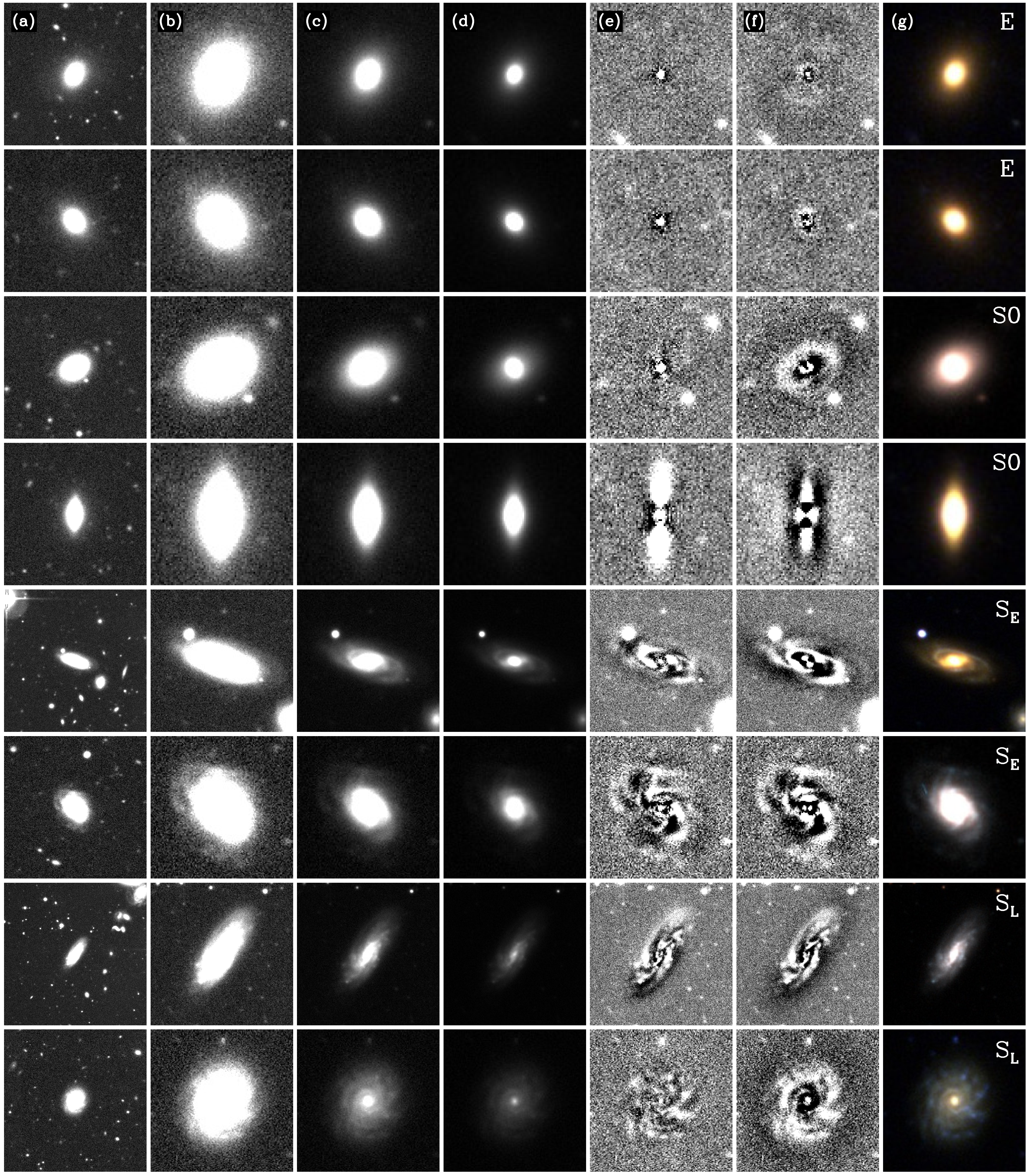}
       \caption[Sample image set for Hubble types]
      {Sample cutout image set for visual inspection. (a) $\rmag$-band cutout image covering eight times $\overline{r_{90}}$. (b)--(d) $\rmag$-band cutout images with different contrasts covering four times $\overline{r_{90}}$. (e) residual image from ellipse fitting. (f) residual image from {\sc {galfit}} model. (g) 2 ($\gmag$ and $\rmag$) or 3 ($\umag$, $\gmag$, and $\rmag$) color-composite image. The morphological class based on the Hubble classification is shown in the top-right corner of panel (g).}
       \label{appvis}
     \end{figure}

     \begin{figure}
       \centering
       \includegraphics[width=1\textwidth]{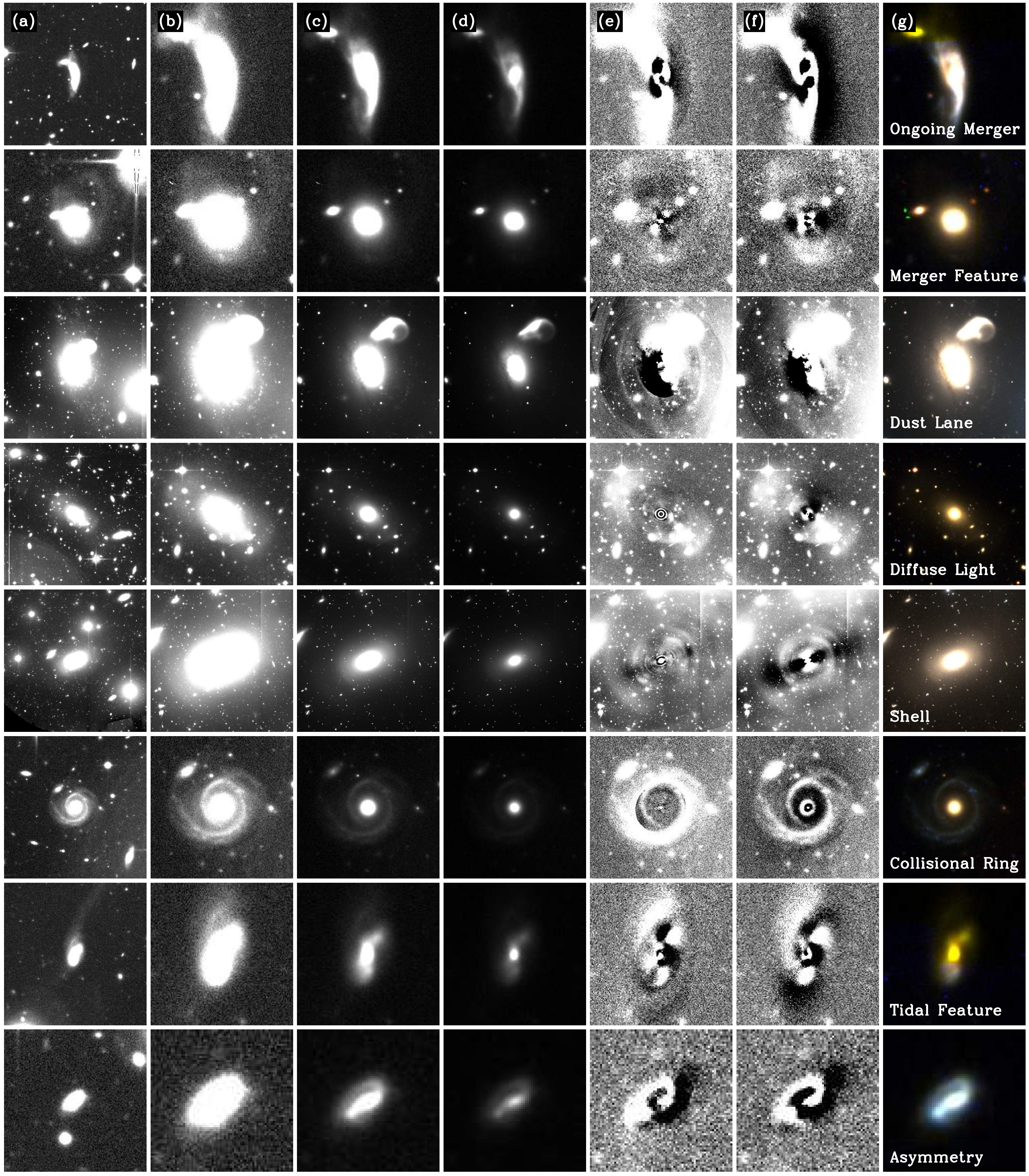}
       \caption[Sample image set for visual inspection]
      {Same as Figure~\ref{appvis} but for galaxies showing ongoing merger feature, post-merger feature, unusual dust lane, asymmetric diffuse light, shell structure, collisional ring, tidal feature, or asymmetric structure.  }
       \label{appmer}
     \end{figure}

\section{The KYDISC catalog}
 \label{app:cat}

 \begin{table*}
   \begin{center}
    \caption{Contents of KYDISC catalog}
    \begin{tabular}{l l l l}
      \hline \hline
      Num & Label & Unit & 	Explanations\\
      \hline    
      1 & ID & - & Galaxy ID \\
      2 & RA & deg & Right accession (J2000) \\
      3 & DEC & deg & Declination (J2000) \\
      4 & $z$ & - & Redshift\\
      5 & $z_{\rm origin}$ & - & The origin of redshift: du Pont=0, IMACS=1, WIYN=2, HeCS=3, NED=4, SDSS=5, and 6dF=6   \\
      6 & $z_{\rm error}$ & - & The error for redshift \\
      7 & $\umag$ & mag & Extinction-corrected $\umag$ magnitude\\
      8 & $\gmag$ & mag & Extinction-corrected $\gmag$ magnitude\\
      9 & $\rmag$ & mag & Extinction-corrected $\rmag$ magnitude\\
      10 & $\umag_{\rm err}$ & mag & Error for $\umag$ magnitude\\
      11 & $\gmag_{\rm err}$ & mag & Error for $\gmag$ magnitude\\
      12 & $\rmag_{\rm err}$ & mag & Error for $\rmag$ magnitude\\
      13 & $\umag_{\rm kcor}$ & mag & $k-$correction for $\umag$ magnitude\\
      14 & $\gmag_{\rm kcor}$ & mag & $k-$correction for $\gmag$ magnitude\\
      15 & $\rmag_{\rm kcor}$ & mag & $k-$correction for $\rmag$ magnitude\\      
      16 & $\overline{r_{\rm e}}$ & $^{\prime\prime}$& Circularized effective radius in $\rmag$ band \\
      17 & $\epsilon_{\rm e}$ & - & Ellipticity at $r_{\rm e}$ \\
      18 & PA$_{\rm e}$ & deg & Position angle at $r_{\rm e}$ (N=90, E=0) \\
      19 & $\overline{r_{\rm 90}}$ & $^{\prime\prime}$& Circularized radius containing 90\% of the total light in $\rmag$ band\\      
      20 & $\epsilon_{\rm 90}$ & - & Ellipticity at ${r_{\rm 90}}$ \\
      21 & PA$_{\rm 90}$ & deg & Position angle at ${r_{\rm 90}}$ (N=90, E=0) \\
      22 & R/R$_{200}$ & - & Cluster-centric distance   \\                                                          
      23 & log$\Sigma_{10}$ & - & Local density  \\  
      24 & $B/T$ & - & Bulge-to-total ratio in $\rmag$ band using \sc{galfit} \\  
      25 & $A$ & - & CAS asymmetry parameter in $\rmag$ band   \\  
      26 & T$_{\rm Hubble}$ & - & E=0, S0=1, S$_{\rm E}$=2, and S$_{\rm L}$=3  \\
      27 & $N_{\rm S, Hubble}$ & - & The significance of the Hubble classification  \\
      28 & T$_{\rm Merger}$  & - & Normal=0, OM=1, and PM=2  \\
      29 & $N_{\rm S, Merger}$ & - & The significance of the Merger classification  \\

 \hline
 \end{tabular}
 \label{tab:cat}
 \tablecomments{Only a portion of this table is shown here to demonstrate its form and content. Machine-readable version of the full table are available.}

\end{center}
\end{table*}

\clearpage

\end{document}